\documentclass[sn-mathphys,Numbered]{sn-jnl}% Math and Physical Sciences Reference Style

\usepackage{graphicx}%
\usepackage{multirow}%
\usepackage{amsmath,amssymb,amsfonts}%
\usepackage{amsthm}%
\usepackage{mathrsfs}%
\usepackage[title]{appendix}%
\usepackage{xcolor}%
\usepackage{textcomp}%
\usepackage{manyfoot}%
\usepackage{booktabs}%
\usepackage{algorithm}%
\usepackage{algorithmicx}%
\usepackage{algpseudocode}%
\usepackage{listings}%
\usepackage{amsmath}
\usepackage{amssymb}
\DeclareMathOperator{\sech}{sech}
\begin{document}

\title[Article Title]{Thermodynamics of Dissipative Solitons}

\author*[1]{\fnm{Vladimir L.} \sur{Kalashnikov}}\email{vladimir.kalashikov@ntnu.no}

\author[1]{\fnm{Alexander} \sur{Rudenkov}}\email{alexander.rudenkov@ntnu.no}
\equalcont{These authors contributed equally to this work.}

\author[1,2]{\fnm{Irina T.} \sur{Sorokina}}\email{irina.sorokina@ntnu.no}
\equalcont{These authors contributed equally to this work.}

\affil*[1]{\orgdiv{Department of Physics}, \orgname{Norwegian University of Science and Technology}, \orgaddress{\street{H$\o$gskoleringen 5}, \city{Trondheim}, \postcode{N-7491}, \country{Norway}}}

\affil[2]{\orgname{ATLA lasers AS}, \orgaddress{\street{Richard Birkelands vei 2B}, \city{Trondheim}, \postcode{N-7034}, \country{Norway}}}

\abstract{We establish a close analogy between the thermodynamics of the nonlinear systems far from equilibrium and the dissipative solitons. Unlike the solitons in the Hamiltonian systems, their dissipative counterpart looks like an aggregation of bounded quasi-particles interacting on the short range, obeying the Rayleigh-Jeans distribution, and possessing a temperature, entropy, and other thermodynamic characteristics. This ensemble is confined by a collective potential, which defines its negative chemical potential. Such a dissipative soliton represents a strongly chirped pulse generated by a mode-locked laser with the advantage of being energy scalable by the analogy with the Bose-Einstein condensation from an incoherent ``basin.'' We demonstrate the main limits of the dissipative soliton energy scaling which results from the loss of internal soliton coherency and the thermalization due to nontriviality of a ``free energy landscape.}




\maketitle
\section{Introduction} \label{sec1}

Dissipative solitons (DS), particularly chirped dissipative solitons in lasers \cite{grelu2012dissipative}, have attracted significant research interest due to their intriguing properties and potential applications in the physics of nonlinear phenomena far from equilibrium \cite{Akhmediev2005,ankiewicz2008dissipative,purwins2010dissipative,ferreira2022dissipative}. The concept of DS encompasses various physical phenomena, including dissipative soliton resonance (DSR) in lasers \cite{akhmediev2009dissipative,grelu2010dissipative}, and Bose-Einstein condensates (BEC) \cite{malomed2005dissipative,fisher2009collective}, astrophysics and cosmology \cite{chavanis2017dissipative,parker1993relativistic}, dynamical processes in complex socio-technical systems \cite{vespignani2012modelling}, and transition to turbulence \cite{kalashnikov2016optics} interweaving the concepts of the partially coherent solitons and week turbulence \cite{picozzi2014optical}, statistical mechanics and thermodynamics \cite{kalashnikov2023dissipative}.

The statistical mechanics and thermodynamics of multi-mode dynamics in fiber optics have been the subject of extensive research. In particular, the interplay between nonlinear effects and multiple interacting spatial modes gives rise to rich phenomena that can be explored using statistical and thermodynamic approaches \cite{churkin2015wave,baudin2023observation}. Recently, a thermodynamic theory of highly multi-mode nonlinear optical systems has been conjectured \cite{wu2019thermodynamic}. This study explores the thermodynamics of complex systems with numerous interacting modes and is closely connected with the phenomena of soliton incoherence and turbulence \cite{zakharov2004one,picozzi2009thermalization,picozzi2014optical}. 

The promising insight into the kinetic theory and thermodynamics of solitons appeared when the soliton spectra were connected to a thermalized Rayleigh-Jeans distribution characterizing turbulence (e.g., see \cite{dyachenko1992optical,robinson1997nonlinear,picozzi2007towards,nazarenko2011wave,baudin2023observation,picozzi2014optical}), that reveals an internal connection between ordered (solitonic) and chaotic (collapsing) states in nonlinear systems \cite{dyachenko1992optical,robinson1997nonlinear}. The exploration of partially coherent solitons as an ensemble of quasi-particles has offered valuable insights into their statistical behavior and the emergent thermodynamic properties inherent in these systems \cite{marcuvitz1980quasiparticle,akhmediev1998partially,picozzi2007towards}. This approach allowed connecting the process of the optical soliton formation and its properties with the physics of BEC \cite{sob2013bose,kalashnikov2021metaphorical} both in the weak \cite{picozzi2014optical} and strong \cite{carusotto2013quantum} nonlinear regimes. 

The fruitfulness of the statistical and thermodynamic viewpoints was demonstrated in the exploration of the DS self-emergence (or mode-locking self-start), which was interpreted as the first-order phase transition or the noise-induced escape from a meta-stable state \cite{gordon2003inhibition,gat2004solution,gordon2006self}. An entanglement of the quantum noise as a ``basin'' within which a soliton self-emergent \cite{gordon2002phase} and its internal structure (e.g., \cite{werner1997phase}) could have a fundamental impact on developing the quantum theory of DS (for a short overview, see \cite{kalashnikov2016optics}).  

The understanding and application of statistical and thermodynamic concepts to DS
are still quickly evolving areas of research, and further investigations may shed light on the potential connections and implications of these concepts to DS. In particular, the energy scaling of DS could be limited by a transition from stable DS to turbulent states \cite{kalashnikov2018self}. In this article, we expose shortly the adiabatic theory of DS, which demonstrates its close connection with the thermodynamics of incoherent solitons and turbulence. The principles of DS formation and DSR are reviewed, and the problem of DS self-emergence from quantum noise is analyzed. The behavior of the DS thermodynamic characteristics with the energy scaling and the soliton thermalization are pointed as the sources limiting the DS energy scalability.

\section{Adiabatic theory of dissipative solitons}\label{sec1}

In contrast to the classical soliton of the nonlinear Schr{\"o}dinger equation, which has a vast field of application ranging from biology and meteorology to photonics and field theory \cite{malomed1}, DS has a nontrivial internal structure \cite{ankiewicz2008dissipative}. Such a structure caused by internal energy flows leads to a phase inhomogeneity characterized by such a characteristic as a ``chirp'' $\Psi=\frac{\partial^2 \phi(t)}{\partial t^2}$. Here, $\phi(t)$ is a phase, and $t$ is a ``local time'', i.e., a time in the system coordinates comoving with DS\footnote{This definition corresponds to photonics, for instance. For BEC, this corresponds to one of the spatial coordinates \cite{kalashnikov2021metaphorical}.}. The $\Psi$ value provides a DS energy scalability (or a mass condensation for BEC) \cite{grelu2012dissipative}. An orientation on the strongly-chirped DS allows the development of their adiabatic theory, the essence of which is of fellows.

\subsection{DS parametric space and dissipative soliton resonance}\label{sub1}

We will base our analysis on the complex nonlinear cubic-quintic Ginzburg-Landau equation (CQGLE), which is a standard model for the study of a broad class of nonlinear phenomena far from equilibrium \cite{aranson2002world,ankiewicz2008dissipative,ferreira2022dissipative,podivilov2005heavily}: 

\begin{gather}\label{eq:CQGLE}
 \frac{\partial}{\partial z} a\! \left(z,t\right)=-\sigma a\! \left(z,t\right)+\left(\alpha+i \beta\right) \frac{\partial^{2}}{\partial t^{2}} a\! \left(z,t\right)-\nonumber \\
 -i \gamma P\! \left(z,t\right) a\! \left(z,t\right)+\kappa\left(1-\zeta P\! \left(z,t\right)\right) P\! \left(z,t\right) a\! \left(z,t\right).   
\end{gather}

\noindent Here, $z$ and $t$ are the propagation coordinates and local time, respectively (see \cite{kalashnikov2021metaphorical} for a comparison with the Gross-Pitaevskii equation). $a(z,t)$ is a slowly-varying field envelope, and $P(z,t)=|a(z,t)|^2$ is a power. The parameters are: $\sigma$ is a saturable net gain, which consists of saturable gain minus loss coefficients; $\alpha$ is an inverse squared bandwidth of a spectral filter; $\beta$ is a group-delay dispersion (GDD) coefficient; $\gamma$ is a self-phase modulation (SPM) coefficient; $\kappa$ and $\zeta$ describe a saturable nonlinear gain (self-amplitude modulation, SAM).    

Below, we will consider the strongly chirped DS with the dimensionless chirp parameter $\psi \propto \gamma^2/\kappa \chi \gg 1$. This requires satisfying the following \textbf{Proposition I}: $\beta/\alpha, \gamma/\kappa \ll 1$, that is, the nondissipative factors dominate the dissipative ones. Since the large chirp means a fast phase change with $t$, we can use the \textbf{Proposition II (adiabatic approximation)}: $\frac{\partial^2 \sqrt{P}}{\partial t^2} \ll 1$, that means a slow change of the DS envelope in comparison with the phase change.  At last, the \textbf{Proposition III}: $C=\alpha \gamma/\kappa \zeta \simeq 1$ will be used. It corresponds proximity to the soliton or potential condition required for the Gibbs-like statistics \cite{katz2006non} and DSR \cite{chang2008dissipative}.

The stationary solution ansatz $a\! \left(z,t\right)=\sqrt{P\! \left(t\right)}\, {\mathrm e}^{i \phi\left(t\right)-i q z}$ ($q$ is a propagation constant or wave-number) reduces Eq. (\ref{eq:CQGLE}) to ($\Omega(t)=\frac{d \phi(t)}{dt}$):

\begin{gather}\label{eq:eq3}
    2\beta P\! \left(t\right) \frac{d^{2}}{d t^{2}}P\! \left(t\right) -\beta\left(\frac{d}{d t}P\! \left(t\right)\right)^{2}  +4 \alpha P\! \left(t\right)\Omega \! \left(t\right)\frac{d}{d t}P\! \left(t\right) +\nonumber \\
    +4 P\! \left(t\right)^{2} \left(-\beta  \Omega \! \left(t\right)^{2}+\alpha  \frac{d}{d t}\Omega \! \left(t\right)-\gamma  P\! \left(t\right)+q\right)=0, \label{eq:eq2}\\
    2 \alpha P\! \left(t\right) \frac{d^{2}}{d t^{2}}P\! \left(t\right)  -\alpha \left(\frac{d}{d t}P\! \left(t\right)\right)^{2}  -4\beta P\! \left(t\right) \Omega \! \left(t\right)  \frac{d}{d t}P\! \left(t\right)  -\nonumber \\
    -4 P\! \left(t\right)^{2} \left(\kappa  \zeta  P\! \left(t\right)^{2}+\alpha  \Omega \! \left(t\right)^{2}+\beta\frac{d}{d t}\Omega \! \left(t\right)  -\kappa  P\! \left(t\right)+\sigma \right)=0, 
\end{gather}

\noindent that, after using the first and second propositions and some algebra, leads to \cite{podivilov2005heavily,kharenko2011highly,kalashnikov2016optics}:

\begin{equation}\label{eq:eq4}
       P\! \left(t\right)=-\frac{\beta \Omega\! \left(t\right)^{2}-q}{\gamma}, 
       \end{equation}
       \begin{equation}\label{eq:eq5}
            \frac{d}{d t} \Omega\! \left(t\right)=\frac{1}{3} \frac{\beta \kappa \zeta \left(\Delta_{\pm}^{2}-\Omega\! \left(t\right)^{2}\right) \left(\Xi_{\pm}^{2}+\Omega\! \left(t\right)^{2}\right)}{\gamma^{2}}, 
       \end{equation}

\noindent where, 

\begin{equation} \label{eq:eq6}
    \Delta_{\pm }^2=\frac{\gamma P_{\pm} }{\beta},
    \end{equation}
    \begin{equation}\label{eq:eq7}
        \beta \Xi_{\pm}^{2}=\frac{\gamma}{\zeta}\left(1 + C -\frac{5}{3} \zeta P_{\pm} \right),
    \end{equation}
   \begin{equation}\label{eq:eq8}
       P_{\pm}=\frac{3}{4} \frac{1-\frac{C}{2} \pm\sqrt{\left(1-\frac{C}{2} \right)^{2}-\frac{4 \zeta \sigma}{\kappa}}}{\zeta}.
   \end{equation}

\noindent Here, $\Delta_{\pm }$ is a cut-off frequency: $\Omega(t)^2\leqslant \Delta_{\pm}^2$; $\Xi_{\pm}$ is a ``chemical potential'' whose meaning will be made clear below, and $P_{\pm}$ is a DS peak power. The $\pm$-signs correspond to the \textit{two branches} of DS solutions.

The appearance of the cut-off frequency $\Delta_{\pm}$ follows from Eq. (\ref{eq:eq4}): $P(t)\ge0$ by definition, therefore $\Omega(t)^2 \le \Delta_{\pm}^2 = q/\beta$. Since $\Omega(0)=0$ by definition, the DS wavenumber is of $q = \beta \Delta_{\pm}^2 =\gamma P_{\pm}$, which has the same sign and is half of that for the Schr{\"o}dinger soliton (SS) \cite{akhmanov1992optics}. Since the sign of the GDD term in Eq. (\ref{eq:CQGLE}) is negative for the nonlinear Schr{\"o}dinger equation, SS does not interact with the linear waves, which have the wavenumber defined by GDD\footnote{The linear waves obey the Bogoliubov (or Langmuir) dispersion relation  \cite{nazarenko2011wave,laurie2012one}.}: $k_l=\beta \omega_l^2 \le 0 \neq q_{SS}=\beta P(0)/2$. However, there is a \textit{resonance with the linear waves} for DS: $q = k_l$, which defines the cut-off frequency $\Delta_{\pm} = q/\beta$ \cite{sorokin2013chaotic}. Thus, the existence of this dispersive resonance could be considered a key factor for DS formation (see Subsection 2.2).

Eq. (\ref{eq:eq5}) results from a factorization procedure aimed to avoid a singularity $\frac{d \Omega(t)}{dt}\to \infty$ (see \cite{maple} for details). This equation brings us to a spectral domain that is an advance of the theory considered. The assumption $\psi \gg 1$ allows using the stationary phase approximation for the Fourier image of the DS amplitude:

\begin{gather}\label{eq:eq9}
    e\! \left(\omega\right)=\sqrt{\frac{\beta}{\gamma}}\int_{-\infty}^{\infty}{\mathrm e}^{-i \omega t} \sqrt{\left(\Delta_{\pm}^{2}-\omega^{2}\right)}\, {\mathrm e}^{\mathrm{I} \phi\left(t\right)}d t \approx \\
   \approx  e(\omega)=\sqrt{\frac{6\pi\gamma}{\zeta  \kappa}} \frac{{\mathrm e}^{\frac{3 \gamma^{2}}{2 \beta  \kappa  \zeta}\frac{ i  \omega^{2}}{  \left(\Xi_{\pm}^{2}+\omega^{2}\right) \left(\Delta_{\pm}^2 -\omega^2 \right)}} }{\sqrt{  i\left(\Xi_{\pm}^{2}+\omega^{2}\right) }}\mathcal{H}\left(\Delta_{\pm}^2-\omega^2\right). \nonumber 
\end{gather}

Here, $\mathcal{H}(x)$ is a Heaviside function, and the proportionality of the chirp $\Psi \propto \gamma^2/\beta \kappa \zeta$ is visible. However, this chirp is inhomogeneous and cannot be compensated perfectly by some second-order GDD of the opposite sign for the DS compression without unavoidable energy loss. Maximum fidelity (or chirp homogeneity) of such compression corresponds to the condition $\Delta_{\pm}^2 = \Xi_{\pm}^2$ \cite{podivilov2005heavily,zhu2013generation}.

Eq. (\ref{eq:eq9}) leads to two important conclusions: 1) \textbf{DS energy} can be expressed as

\begin{equation} \label{eq:eq10}
   E= \frac{1}{2\pi}\int_{-\infty }^{\infty }p(\omega)d\omega=\frac{6 \gamma \arctan\! \left(\frac{\Delta_{\pm }}{\Xi_{\pm }}\right)}{\zeta \kappa \Xi_{\pm }},
\end{equation}
\noindent where 2) \textbf{DS power} $p(\omega)$ is

\begin{equation}\label{eq:eq11}
    p(\omega)=\left| e(\omega) \right|^2=\frac{6 \pi \gamma \mathcal{H}\! \left(\Delta_{\pm}^{2}-\omega^{2}\right)}{\zeta \kappa \left(\Xi_{\pm}^{2}+\omega^{2}\right)}. 
\end{equation}

Eq. (\ref{eq:eq10}) allows constructing the \textit{master diagram} representing the two-dimensional DS parametric space $C$ vs. $E$ \cite{podivilov2005heavily,rudenkov2023high,maple}. Such a diagram shows the stability threshold (maximal) $C(E)|_{\sigma=0}$, which is connected with the region of the $+$-solution. $\sigma$ is positive above this threshold, i.e., a vacuum is unstable for a larger $C$, which value varies from 2 to 2/3 with energy. The next important component is a $\pm$ border: the $-$-solution lies below $+$ one on the $C$-parameter. The maximal fidelity curve on the master diagram corresponds to the condition of $\Delta_{+}=\Xi_{+}$. At last, the master diagram demonstrates a network of the isogains $\Sigma=\frac{4 \zeta \sigma}{\kappa}=const$.

One can see that the DS spectrum has a shape of the Lorentzian function of the $\Xi_{\pm}$- width truncated at $\pm \Delta_{\pm}$. In fact, it reproduces a Rayleigh-Jeans distribution specific to turbulence \cite{robinson1997nonlinear,nazarenko2011wave,picozzi2014optical}. As would be shown below, it is not only a formal analogy.

The following important consequence from the adiabatic theory of DS is a formulation of the \textbf{dissipative soliton resonance} (DSR) conditions \cite{chang2008dissipative}. DSR corresponds to a perfect DS energy scalability. Formally, DSR can be defined as $\exists \,C^*:\lim_{C \to C^*}E=\infty$ or there exists a set of $C$-parameters providing infinite energy asymptotics. The region of DSR belongs to the $+$-branch of DS, which bottom border is defined by 

\begin{equation}\label{eq:eq12}
    \begin{matrix}
E=\frac{6 \sqrt{2\gamma \beta}}{\kappa \sqrt{\eta}}\frac{\arctan(\frac{\sqrt{3}\sqrt[4]{\Sigma}}{\sqrt{6-13\sqrt{\Sigma}}})}{\sqrt{6-13\sqrt{\Sigma}}}, \\
C=2-4\sqrt{\Sigma}, \\
P_0=\frac{3\sqrt{\Sigma}}{2 \zeta}, \\
\Delta^2= \frac{3\gamma\sqrt{\Sigma}}{2\beta \zeta},\\
\Xi^2=\frac{\gamma}{2\zeta \beta}(6-13\sqrt{\Sigma}),
\end{matrix}
\end{equation}
\noindent so that DSR exists within $\Sigma_{+} \in \left[ 0,36/169 \right]$ and $C_{+} \in \left[ 2/3, \,2/13\right]$. The main \textit{visible signatures of the transition to DSR} are: 1) $\Xi_{+} < \Delta_{+}$, 2) $\Delta_{+} \to const$, 3) $P_{+} \to const$. The energy scaling proceeds through the DS width and chirp scaling. 

The limiting case of DSR corresponding to $\Sigma \to 0$ gives $\Delta_{+}^2=\gamma/ \beta \zeta$, $P_{+}=\zeta^{-1}$, and $\Xi_{+}=0$. However, the stored energy in a laser cavity with a period $T_{cav}$ is confined by $(\zeta T_{cav})^{-1}$. This defines a minimum value of $\Xi_{+}$:

\begin{equation} \label{eq:eq13}
    1=\frac{6 \gamma}{\kappa T_{cav}\Xi_{+}}\arctan{\left( \frac{\sqrt{\frac{2\kappa}{3\zeta}}}{\Xi_{+} \sqrt{\alpha}} \right)}.
\end{equation}

Eq. (\ref{eq:eq13}) gives a minimum of $\Xi_{+}$ and demonstrates a balance of two dimensionless scales: $\Xi_{+} T_{cav}$ and $\Xi_{+} \sqrt{\alpha}$. Also, one should take into account that $T_{cav}$ is confined above by a value of the gain relaxation time $T_r$ \cite{kalashnikov2005approaching}, not to mention the fact that a laser average power $\sim \zeta^{-1}$ is unreachable\footnote{For instance, these parameters equal approximately 4.9 $\mu$s and 30 kW for a Cr:ZnS laser, respectively \cite{rudenkov2023high}.}. These facts put physical limitations on the DS energy scaling.

\subsection{DS without a spectral dissipation}

As was mentioned above, the dissipative resonance $\beta \omega_l^2 = q$, defining the cut-off frequency $\Delta_{\pm}$ is the forming factor for DS. As another forming factor, one may point out the energy balance between the spectral loss and nonlinear gain: $\alpha \Delta_{\pm}^2 \simeq \kappa P_{\pm}$. So that, in combination with the dispersion resonance condition, one has $C=\alpha \gamma/\beta \kappa \simeq 1$ that is close to the DSR condition and corresponds to the Proposition III \cite{kalashnikov2018self}. The importance of spectral filtering as an additional SAM mechanism was pointed out in \cite{bale2008spectral}: the chirp scatters the frequency to the DS front and tail, where they undergo the spectral loss, which confines DS in the temporal domain. However, we do not consider this argument as comprehensive. The point is that Eq. (1) has the DS solutions for $\alpha=0$: spike on constant background and tabletop or truncated spike \cite{maple2}. Repeating the approach based on the adiabatic theory gives the truncated convex spectrum similar to that in \cite{PhysRevA.79.043829}:

\begin{equation} \label{eq:eq14}
    p\! \left(\omega \right)=\frac{6 \pi  \beta}{\kappa}  \left| \frac{\Delta^{2}-\omega^{2}}{\left(\Xi^{2}+\omega^{2}\right) \left(\frac{3 \Delta^{2} \zeta  \beta}{\gamma}-1\right)}\right|,\ \ \ \omega^2<\Delta^2,
\end{equation}
\noindent and the DS parameters are:

\begin{gather}\label{eq:eq15}
    P_{\pm}=\frac{2 \left(1+\sqrt{-\Sigma +1}\right)}{3 \zeta},\nonumber\\
    \beta \Delta_{\pm}^2 = \gamma P_{\pm}^2,\\
    \Xi_{\pm}=\frac{6 \sqrt{1-\Sigma}\pm7 \Sigma \mp6}{\pm 9+18 \sqrt{1-\Sigma}}\nonumber 
\end{gather}

One should note that the spectrum (\ref{eq:eq14}) is not Rayleigh-Jeans-like and, thus, is not thermalized \cite{picozzi2009thermalization}. One may conjecture that this exotic regime could shed light on the mechanisms of DS formation that need numerical exploration. For this goal, we solved Eq. (1) numerically based on the FFT split-step algorithm with taking into account the net-gain dynamics: $\sigma = \delta (E/E_{cw} -1)$ in (1) ($\delta$ is a ``stiffness'' parameter, $E_{cw}$ is continuous-wave energy, see \cite{kalashnikov2006chirped}). 

We considered a time window covered by $N = 2^{16}$-cells with the  $\Delta t = 10$ fs time-cell size. Eq. (1) was supplemented by an additive complex Gaussian white noise term $\Gamma(z,m \Delta t)$ ($m \in  [1...N]$ is a cell number)\footnote{Here and below, $\Delta t$ and $\Delta \omega$ mean the cell sizes in the temporal and spectral domains, respectively.} with the covariance \cite{206583}:

\begin{gather} \label{eq:eq16}
\left\langle \Gamma_m (z)\Gamma_n^* (z') \right\rangle = \mathcal{W} \delta_{mn}\delta(z-z') \nonumber \\ 
\left\langle \Gamma_m (z)\Gamma_n (z') \right\rangle = 0,\\
\mathcal{W} = 2 \gamma h \nu |\sigma|/T_{cav}.\nonumber 
\end{gather} 

\noindent Here $\mathcal{W}$ is a normalized noise power and $\nu$ is a carrier frequency. $\mathcal{W}$ can be associated with a ``bath temperature''\footnote{This concept of an added noise has no to be missed with the noise temperature of optical amplifier: $T_n=\frac{h \nu}{k_B}\left[ \log(1+\frac{1-e^{-h\nu/k_BT}}{1-|\sigma|^{-2}}) \right]^{-1}$ \cite{gardiner2004quantum}.} \cite{gordon2002phase}. The Euler–Maruyama method was used for the solution of (Eq. (1) + $\Gamma(z,t)$). The initial condition is the noise (16) supplemented by a \textit{sech}-like spike with an amplitude of doubled significant noise high and the width of $10\Delta t$.

The numerical temporal and spectral profiles of DS in the absence of spectral dissipation (i.e., $\alpha=0$ in Eq. (1)) are shown in Fig. 1. The numerical spectra and temporal profiles demonstrate the characteristic signatures of DSR: the flat-top shape of DS and the finger-like spectrum $\Xi^2 \ll \Delta^2$. The spectral peculiarities are i) a pronounced spectral condensation around $\omega=0$ and ii) an energy transport to the higher frequencies which can result in an appearance of the modulated spectral pedestal (Fig. 1,b). The last corresponds to an appearance of perturbation spikes on the DS edges (Fig. 1, a). 

\begin{figure}[h]  \label{fig:fig1}
    \centering
    \includegraphics[width=1\textwidth]{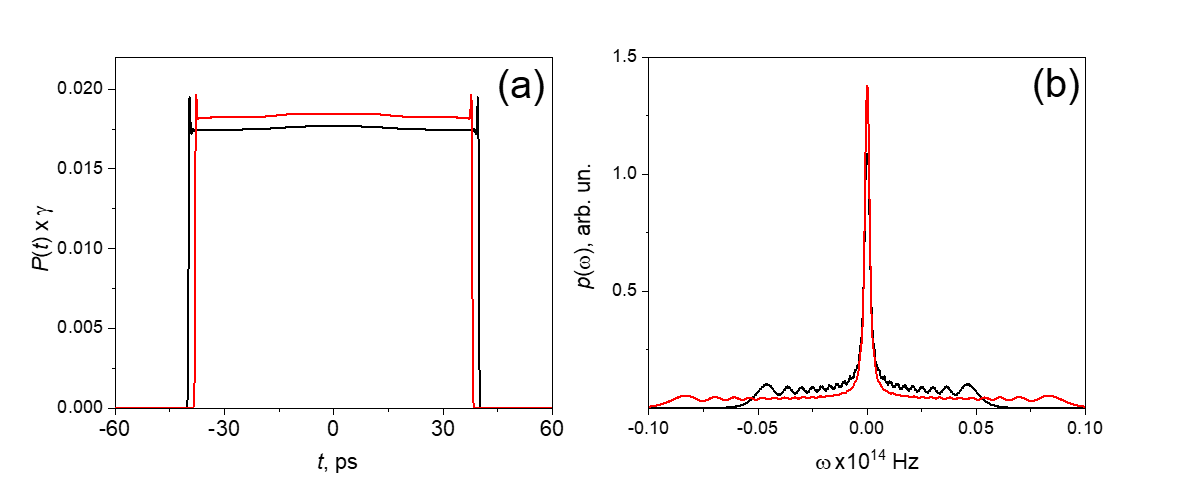}
    \caption{Dimensionless power $P(t)$ (a) and spectral power $p(\omega)$ (b) for DS without (black) and with (red) an additive Gaussian white noise. The parameters correspond to Ref. \cite{robinson1997nonlinear}: $\beta=220$ fs$^2$, $\gamma=5.1$ MW$^{-1}$, $\kappa=4$ MW$^{-1}$, $\kappa=30$ MW$^{-1}$, central wavelength equals 2.27 $\mu$m, and $E_{cw}=150$ nJ. $z=15000$, $\delta=0.04$, and DS energy $E \approx 38$ nJ for the 15\% output coupler. No spectral dissipation ($\alpha=0$).}
  \end{figure}

The noise presence changes the DS temporal profile and the central part of its spectrum slightly, but the energy transport to the spectral wings enhances. Thus, we can conclude that such exotic chirped DS is (quasi-)stable, and the dispersion relation $\beta \Delta^2 = \gamma P(0)^2 = \beta \omega_l^2$ plays a major role in the DS formation. Such a relation describes a ``high of the collective confining potential,'' \cite{hall2002statistical,picozzi2009thermalization} providing a DS temporal and spectral localization and characterized by a minimal correlation scale of DS $\ell \propto |\Delta|^{-1}$. DS spectrum (Fig. 1, b), which is analog to a turbulence spectrum \cite{robinson1997nonlinear,picozzi2009thermalization}\footnote{The internal connection between chaotic and ordered states in the nonlinear systems like those considered here was pointed in a lot of works. As an example, see \cite{during2009breakdown}.}, could be explained by the existence of two cascades in a spectral domain\footnote{Namely, a direct energy cascade to higher frequencies $|\omega|\to |\Delta|$ and an inverse cascade of a ``spectral density'', i.e., $p(\omega)/\Delta\omega$, to $\omega=0$ \cite{zakharov2004one}.} like that in turbulence \cite{nazarenko2011wave,laurie2012one}. Nevertheless, the spectral cut-off due to $\alpha \ne 0$ could stop the spectral out-flowing\footnote{It is a process of so-called ``evaporation cooling'' leading to a spectral condensation at $\omega=0$ like that in BEC \cite{nazarenko2011wave}.} and, thereby, make DS robust \cite{bale2008spectral} through the ``kinetic cooling'' as in a low-dissipative BEC \cite{kalashnikov2021metaphorical}. In this sense, the spectral dissipation can be treated as an additional mechanism of SAM, which ``strengthens'' an effective potential border that was described by as above as an interrelation $\beta \Delta^2 = \gamma P_0 \iff \alpha \Delta^2=\kappa P_0$.

\section{DS self-emergence}\label{ss}

The DS self-emergence is crucial both practically and theoretically. In laser physics, such an emergence from a noisy ``bath'' is named a mode-locking self-start. The different mechanisms for the spontaneous appearance of DS were proposed, in particular, the spontaneous growth of a field fluctuation from the beat note of the free-running spectrum enhanced by the self-induced refractive index grating was proposed \cite{krausz1991self,krausz1993passive}. As was conjectured, the multimode Risken-Nummedal-Graham-Haken instability can result in spontaneous (but unstable) self-mode-locking \cite{lugiato2015nonlinear}, which can be stabilized by the resonant coupling with the phonons \cite{sergeyev2021vector}. Also, the dynamic gain saturation can play an important role in the fluctuation enhancement required for the DS self-emergence \cite{ippen1990self}. Also, the mode-locking self-start could be treated as a dynamic loss of the continuum-wave stability \cite{chen1995self,soto2002continuous}. Nevertheless, we think that the most prospective and general approach to the problem of the DS self-emergence is based on the thermodynamic approach developed in \cite{gordon2002phase,gordon2003phase,gordon2006self}, which considers this phenomenon as a first-order phase transition or escape from a metastable state \cite{lindner2004effects}.

We base our analysis on the numerical simulations of Eq. (\ref{eq:CQGLE}) with the additive complex noise described above. Statistics were gathered on the $\sim$100 independent samplings for each parametric set. The initial condition in the form of a quantum noise (16) supplemented by a seed $\mathcal{W'} \sech{(t/10\Delta t)}$ was chosen, where the seed amplitude $\mathcal{W'}$ corresponds to a standard rogue wave definition, namely a doubled significant wave high \cite{dudley2019rogue}.

As the basic characteristics, we used a mean self-start time $<\tau>$ and its standard deviation $S$. The DS characteristics were characterized by the order parameter $\wp$ %the Kuramoto order parameter:
 \cite{gordon2002phase,gordon2003melting,gat2004solution}:

\begin{equation} \label{eq:order}
    \mathbf{\wp}=\frac{1}{2}\left\langle\sqrt[4]{\frac{\sum_{n=1}^{N}\left| a_n \right|^4}{\left(\sum_{n=1}^{N}\left| a_n \right|^2 \right)^2}}\right\rangle,
\end{equation}

%\begin{equation}
%   \mathbf{r}=\frac{1}{N}\left|\sum_{n=1}^{N} e^{i\phi_n}\right|, \label{eq:kuramoto}
%\end{equation}

%\begin{equation}
%    \mathbf{\wp}=\frac{1}{2}\sqrt[4]{\frac{\sum_{j-k+l-m=0}^{N} a_j a_k^*a_la_m^*}%{\left(\sum_{j=1}^{N}  \left| a_j \right|^2\right)^2}}, \label{eq:order}
%\end{equation}

\noindent where we use the decomposition for a field on the mesh $n\in \left[ 1,N \right]$:
$a(t) \to a_n = |a(t_n)|e^{i\phi_n}$, and the averaging over an ensemble of different samples is assumed. This parameter shows a fraction of the synchronized modes related to the overall mode amount $N$.

Like the definition (\ref{eq:order}), one may define the system order parameter as 

\begin{equation}\label{eq:order2}
    M=\sigma/\delta=E/E_{cw}-1,
\end{equation}

\noindent which shows a level of the DS energy domination over the continuous wave. Such domination results from wave condensation caused by nonlinearity (cubic-quintic terms in Eq. (1)).

The examples of self-start dynamics are shown in Fig. 2. We could emphasize the following peculiarities of such dynamics. i) There are two thresholds - the first one corresponds to an appearance of CW-radiation at an early stage of dynamics, and the second corresponds to a mode-locking self-start at a later stage. ii) The self-start is ``smooth'' for low energies, i.e., transitioning from CW to mode-locking resembles a second-order phase transition (black curve). iii) The energy growth leads to a ``tough'' mode-locking start (first-order phase transition) with an appearance of a Q-switch spike (spikes) (red curve). iv) Further energy scaling leads to double-pulsing because the chosen $C$ does not fit the DSR condition, which requires $C<2/3$ (Eq. (\ref{eq:eq12}). For some statistical samples, the single DS with a noisy background can develop (Fig. \ref{fig:fig3}). Evolution of the $\wp$-parameter for a given $E_{cw}$ is shown by open red circles in Fig. 2.     

\begin{figure}[h] \label{fig:fig2}
    \centering
    \includegraphics[width=0.6\textwidth]{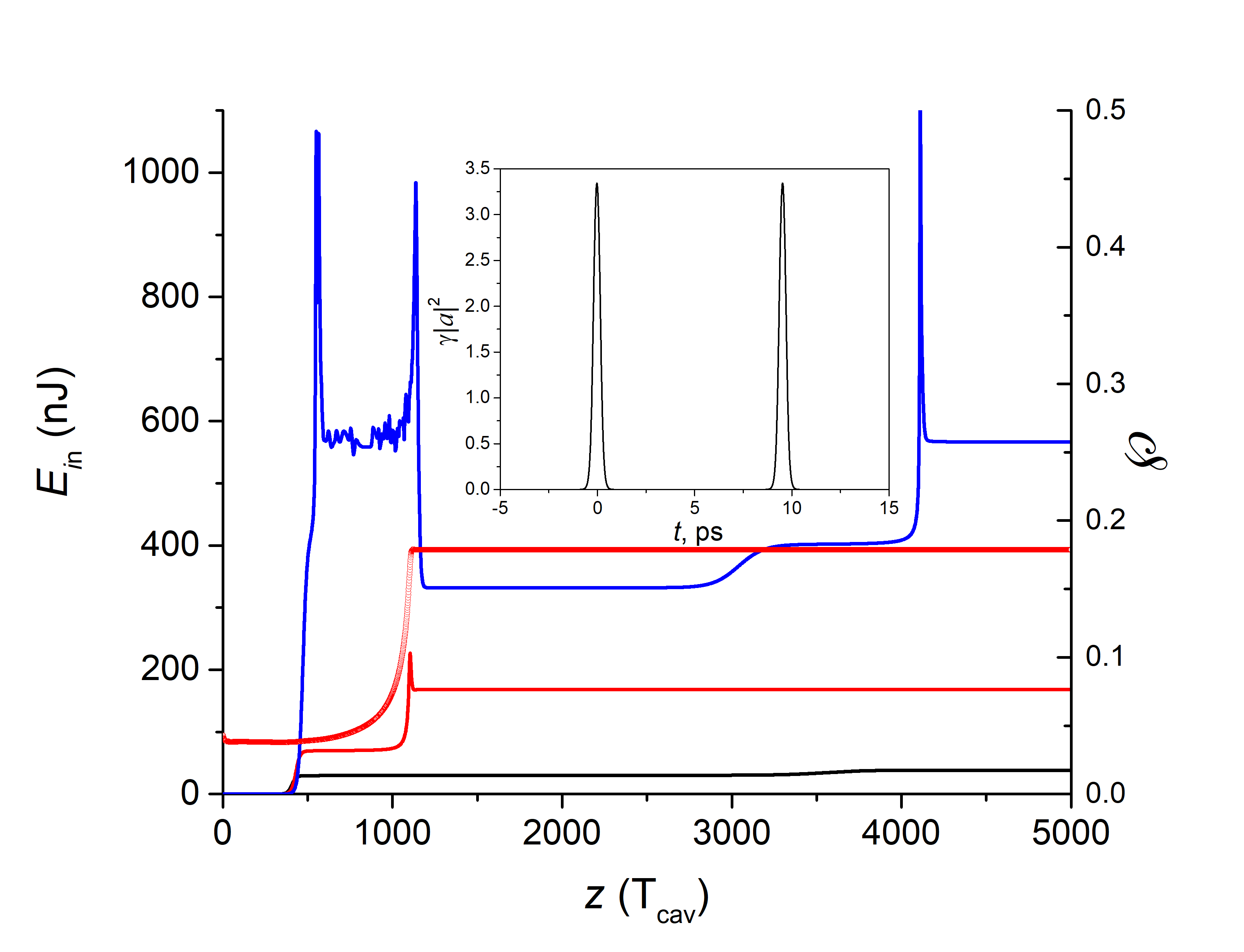}
    \caption{Evolution of the intra-cavity energy $E_{in}$ with $z$. The laser setup corresponds to that described in \cite{rudenkov2023high}: Cr:ZnS active medium with an effective gain bandwidth 200 nm, $\gamma=$5.1 MW$^{-1}$, $\kappa=\zeta=$1 MW$^{-1}$, $\beta=$880 fs$^{2}$, $C=$1.08, and $\delta=0.04$. The curves correspond to $E_{cw}=$ 30 (black curve), 70 (red), and 400 (blue) nJ, respectively, for one of the stochastic samples. Inset shows the dimensional power profile of an intra-cavity field at $z=$5000 $T_{cav}$ for the blue curve. Red open circles show an evolution of the $\wp$-parameter along the red solid curve for $E_{cw}=$ 70 nJ.}
   \end{figure}

\begin{figure}[h] \label{fig:fig3}
    \centering
    \includegraphics[width=1\textwidth]{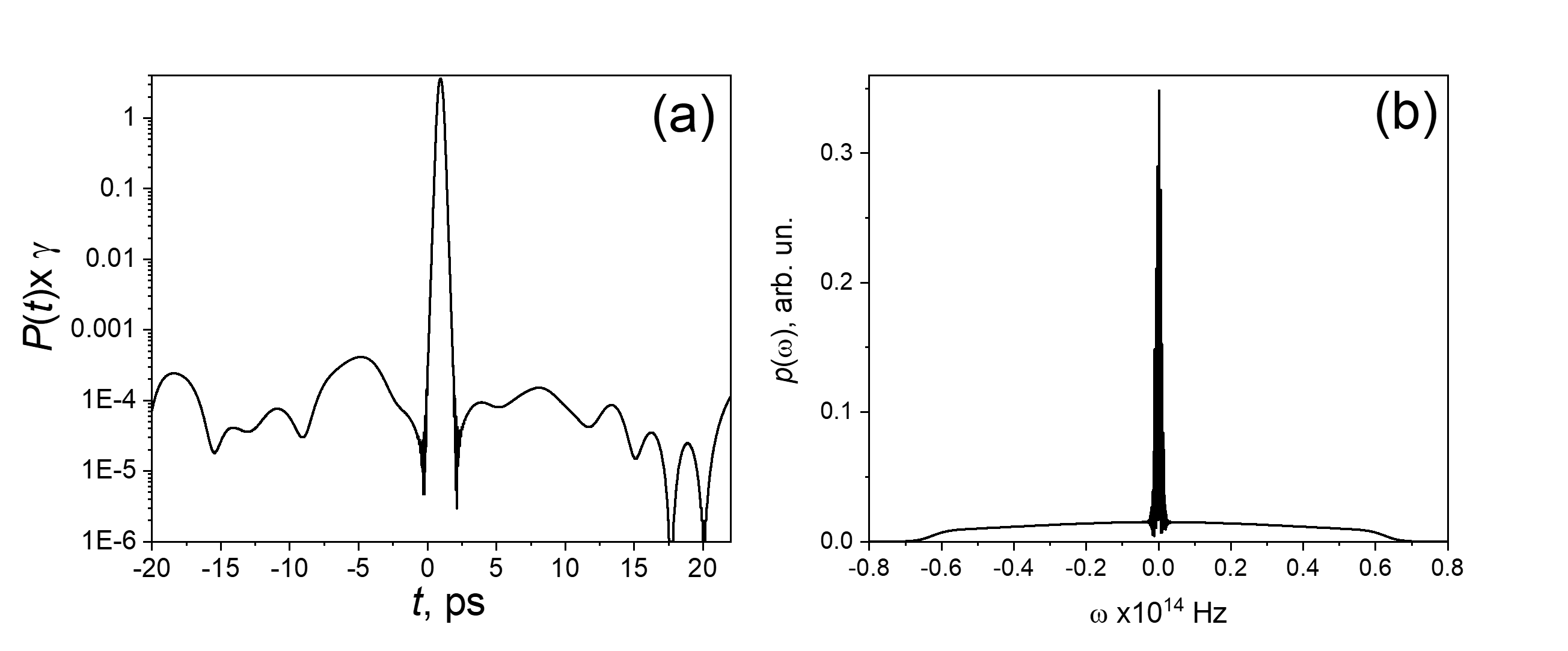}
    \caption{DS power $P(t)$ (a) and spectrum $p(\omega)$ (b) for one of the stochastic samples at $E_{cw}=$350 nJ and other parameters of Fig. 2. The noisy pedestal (a) and corresponding modulated spectrum (b) are clearly visible.}
   \end{figure}

Under DSR conditions (i.e., $2/13<C<2/3$), further energy scaling is possible. Figures 4 demonstrate an energy dependency of $<\tau>$, $S$, $M$, and $\mathbf{\wp}$ as well as the DS temporal ($T$) and spectral ($\Omega$) FHWM widths for a mode-locking regime. Two main regimes corresponding to $``-''$ and $``+''$-branches of DS (see Subsection \ref{sub1}) are clearly visible: the first (low-energy) corresponds to decreasing $T$ and increasing $\Omega$ with the energy growth, the second demonstrates the pulse width stretching and $\Omega$-decrease (i.e., the parameter $\Xi<\Delta$ and decreases) with energy. The division between these regimes is manifested approximately by the maximum order parameters. It is a so-called maximum fidelity point corresponding to the equality of $\Delta=\Xi$, which means the coincidence of the short- and long-range correlation times characterizing DS \cite{rudenkov2023high}. 

\begin{figure}[h]\label{fig:fig4}
    \centering
    \includegraphics[width=1.05\textwidth]{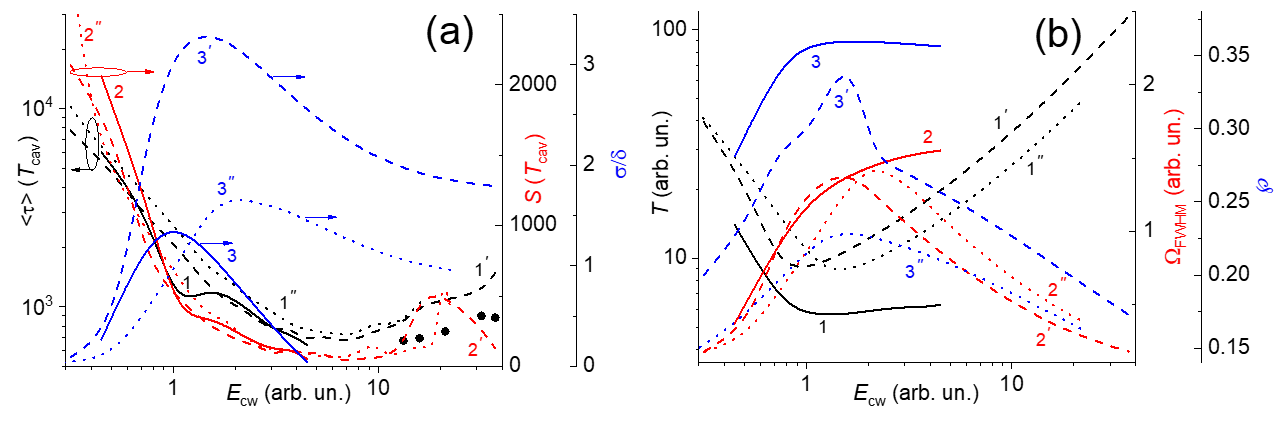}
    \caption{Energy dependencies of (a): averaged DS self-starting time $<\tau>$ (black curves 1), its standard deviations $S$ (red curves 2) and order parameter $\sigma/\delta$ (blue lines 3); (b): DS width $T$ (black lines), FWHM spectral width $\Omega_{FWHM}$ (red lines), and order parameter $\mathbf{\wp}$ (blue lines). Energy is normalized to $\kappa \sqrt{\zeta/\beta \gamma}$, frequency is normalized to $\sqrt{\beta \zeta/\gamma}$, and time is normalized to $\kappa/\sqrt{\beta \zeta \gamma}$. The physical parameters are close to those in Ref. \cite{rudenkov2023high}: $\gamma$=5.1 MW$^{-1}$, $\kappa=$ 1 MW$^{-1}$ (solid curves 1, 2, 3 and dashed curves $1'$, $2'$, $3'$) and 0.5 MW$^{-1}$ (dotted curves $1''$, $2''$, $3''$). $C$=1.08 (solid curves 1, 2, 3) and 0.54 (dashed curves $1'$, $2'$, $3'$ and dotted curves $1''$, $2''$, $3''$). Spectral filter bandwidth equals 200 nm. The energy scaling is performed by the $T_{cav}$ variation assuming $E_{cw}$=150 nJ for $T_{cav}$=81 ns. Points show $<\tau>$ for the dual DSs and $C$=0.54, $\kappa=$ 1 MW$^{-1}$.}
    \end{figure}

The self-starting time $<\tau>$ and the corresponding standard deviation $S$ decrease with energy but then begin to increase (Fig. 4, a). This phenomenon limiting the DS energy scalability will be considered in the next section.

\section{DS thermalization and free energy landscape} \label{therma}

As was pointed out in Subsection \ref{sub1}, the DS spectrum reproduces a Rayleigh-Jeans distribution characterizing an incoherent soliton \cite{picozzi2014optical}. Such a parallel is based on the phase inhomogeneity of strongly chirped DS, which allows treating it as an ensemble of ``quasiparticles'' confined by collective potential \cite{picozzi2007towards}. This treatment leads to the following definitions of thermodynamic characteristics of an \textit{isolated} DS\footnote{``isolated'' in the sense that DS is connected with an environment only through the dissipative ``channels'' defined by Eq. (\ref{eq:CQGLE}) without taking into account a possible dynamics of ``basin'' (``vacuum''), appearance of multiple DSs, etc.}\cite{kalashnikov2023dissipative}.

1) \textbf{DS ``temperature''}:
\begin{equation}
    \Theta=6\pi\gamma/\zeta \kappa,
\end{equation}\label{themperature}
2) negative \textbf{chemical potential}:
\begin{equation}
    -\mu=\Xi^2,
\end{equation}\label{potential}
\backmatter
3) \textbf{Boltzmann entropy}:
\begin{equation}
    S_B = \int_{-\Delta}^{\Delta}\ln{p(\omega)} d\omega,
\end{equation}\label{be}
4) \textbf{Shannon entropy}:
\begin{equation}
    S_{Sh} = -\int_{-\Delta}^{\Delta}p(\omega)\ln{p(\omega)} d\omega,
\end{equation}\label{se}
5) \textbf{internal energy}:
\begin{equation}
    U = \int_{-\Delta}^{\Delta} \omega^2 p(\omega) d\omega,
\end{equation}\label{U}
6) and \textbf{free energy}:
\begin{equation}
    \mathcal{F}=U-\Theta S_B.
\end{equation}\label{free}

Fig. 5 (a) demonstrates the examples of the ``internal'' entropy $S_B$ and free energy calculated from these formulas. One can see that the entropy is minimal in the vicinity of the maximum order parameter and grows with the energy scaling. Also, it increases with ``temperature'' $\Theta \propto 1/\kappa$ (dotted black curve $1''$). Simultaneously, the chemical potential $\Xi$ decreases (see the decreasing $\Omega_{FWHM}$ in Fig. 4 (b)) so that negative free energy tends to $(2\Delta - S_B) \Theta$. All these factors testify that DS diminishes its ability to concentrate energy with the $E_{cw}$-growth. This could limit energy scaling due to transit to turbulence \cite{kalashnikov2023dissipative}. 

\begin{figure}[h] \label{fig:fig5}
    \centering
    \includegraphics[width=1\textwidth]{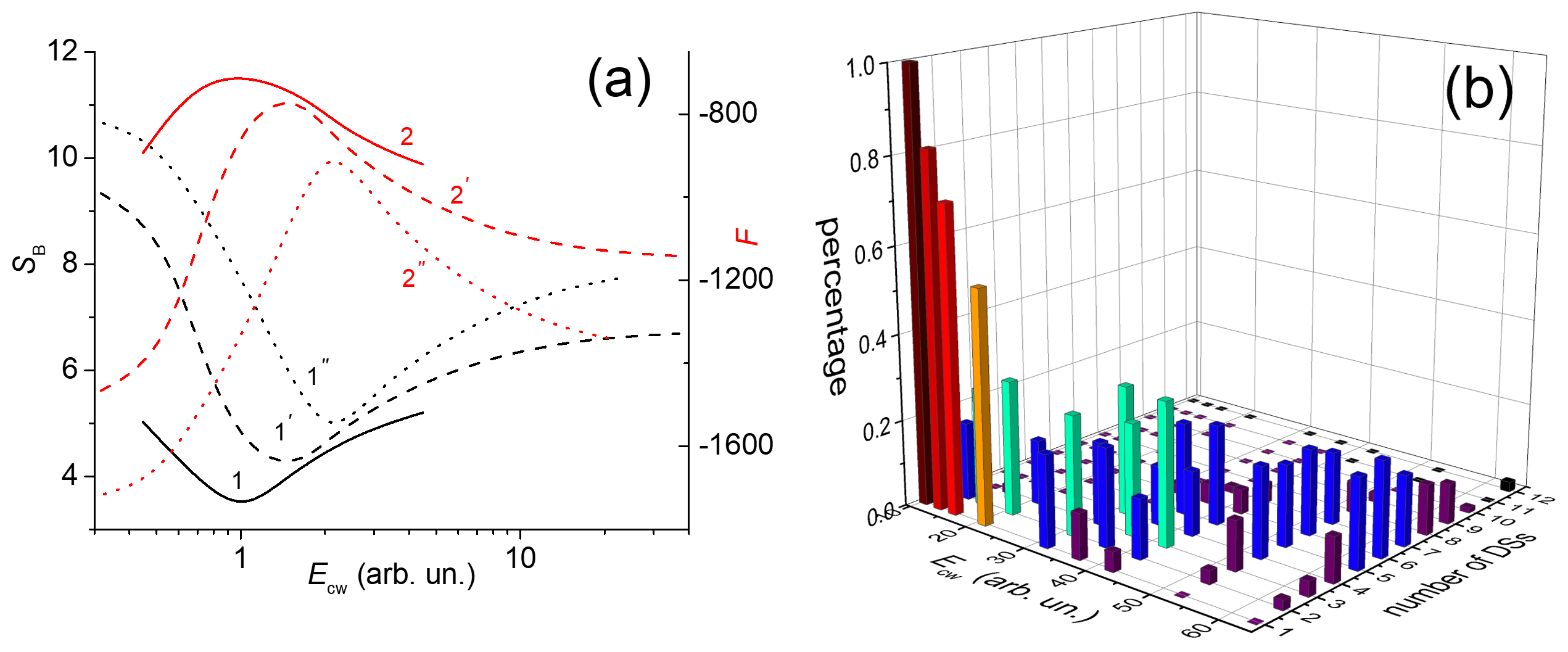}
    \caption{(a): Entropy dependence on the normalized energy (black curves) for the parameters of the black curves in Fig. 4. red curves -- the same for a free energy dependence. (b): Percentage of multiple DSs in an ensemble of 100 stochastic samples (the parameters are the same as in (a)).}
    \end{figure}

However, our analysis demonstrates that the energy scaling is limited by a more strong factor, namely thermalization. If we consider that DS emerges from an initial quantum noise (``vacuum fluctuations''), the excitation of the latter could play a decisive role. That is, if the vacuum is sufficiently ``hot'' (i.e., $E_{cw}$ is large), multiple DSs can emerge in the different stochastic samples. Fig.5 (b) illustrates this situation: after some maximum energy, multiple DSs begin to develop, and their number and contribution increase with energy, which leads to an energy thermalization in a stochastic DS ensemble. It should be noted that DS in some concrete stochastic samples is dynamically stable.

Qualitatively, this phenomenon could be explained by treating the DS self-emergence as a noise-induced escape from a metastable state \cite{hanggi1986escape,gordon2006self}. From this point of view, the DS self-emergence corresponds to transit from a metastable state corresponding to CW-generation (see Fig. 2) to a ``deeper'' minimum of free energy corresponding to DS. In reality, a multitude of such local minima corresponding to different numbers of DS can co-exist so that a system could evolve to one of them stochastically, and an escaping rate would depend on the potential barrier high and a noise temperature (which is $\propto E_{cw}$ in our case). Black points in Fig. 4 (a) demonstrate the growth of an escaping time (i.e., a decrease of the DS self-building time) so that the multi-soliton regimes begin to prevail over the one-soliton operation. In particular, a nontrivial ``geography'' of the free energy landscape is illustrated by a spontaneous switching between DS and SS in a laser system with the higher-order group-delay dispersions (i.e., an $\omega$-dependent $\beta$) illustrated by Fig. 6. The thorough analysis of this conjecture is waiting for further analysis.   

\begin{figure}[h]\label{fig:fig6}
    \centering
    \includegraphics[width=0.6\textwidth]{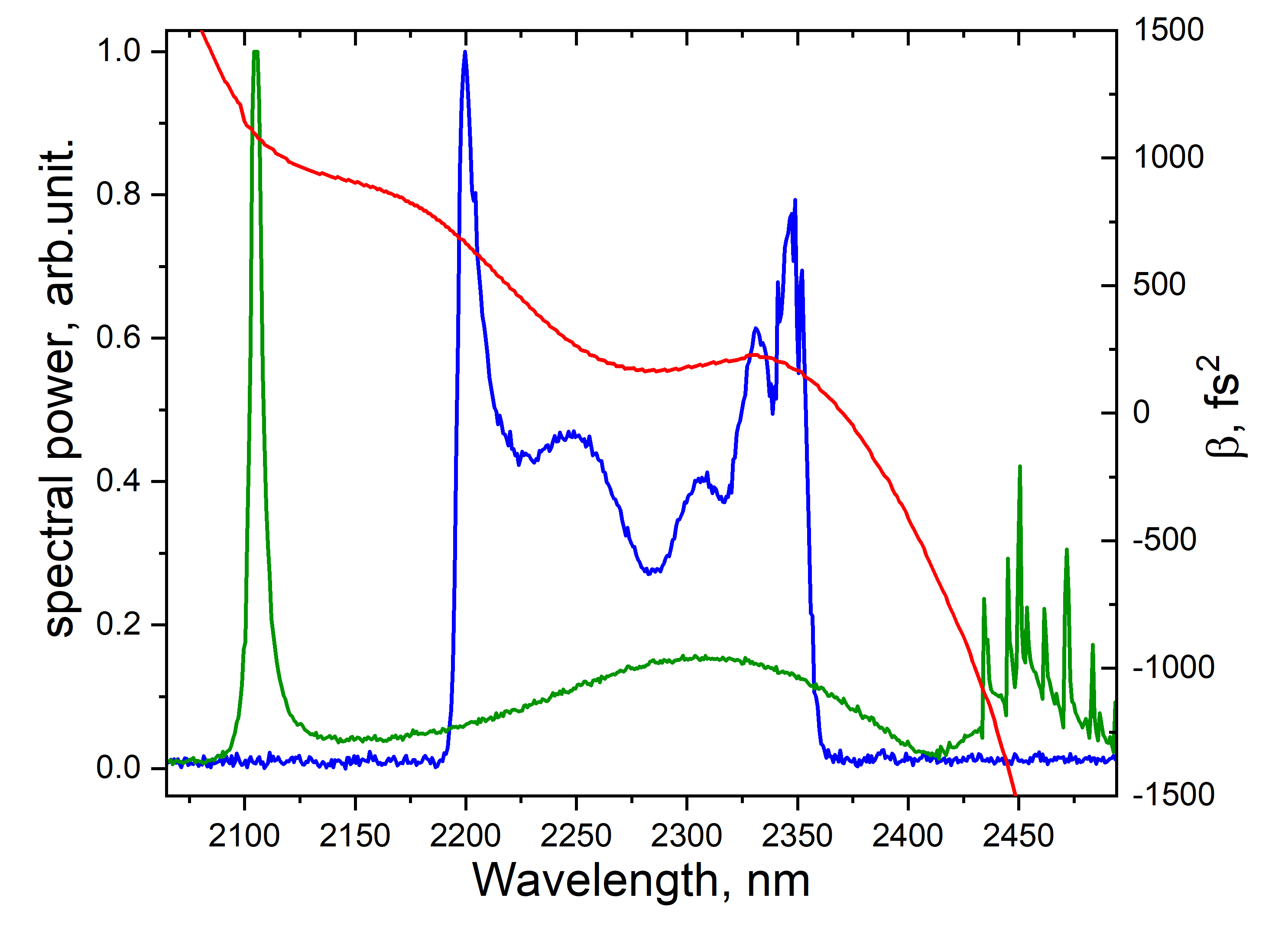}
    \caption{A Cr:ZnS laser described in Ref. \cite{rudenkov2023high} can spontaneously switch between two regimes: DS (spectrum is shown by blue curve) and SS (green curve). Such a nontrivial regime could be explained by frequency-dependent dispersion (red curve).}
    \end{figure}

\section{Conclusion}

The adiabatic theory of DS provides us with a deep inside into the DS properties and mechanisms of its formation. In particular, it demonstrates that these mechanisms resemble those for incoherent solitons and turbulence. That allows applying an ideology of statistical mechanics and thermodynamics for exploring DS and, in particular, defining the limits of its energy scalability or, in other words, the DSR existence. We see two main such factors: i) the growth of internal DS entropy in parallel with the vanishing of its chemical potential and ii) the nontriviality of the free energy landscape, which leads to the DS thermalization, i.e., multiple DS generation.    

\bmhead{Acknowledgments}

The work is supported by the Norwegian Research Council projects \#303347 (UNLOCK), \#326503 (MIR), and by ATLA Lasers AS. We acknowledge using the IDUN cluster \cite{sjalander}.

\bibliography{sn-bibliography}% common bib file


\begin{thebibliography}{69}
\ifx \bisbn   \undefined \def \bisbn  #1{ISBN #1}\fi
\ifx \binits  \undefined \def \binits#1{#1}\fi
\ifx \bauthor  \undefined \def \bauthor#1{#1}\fi
\ifx \batitle  \undefined \def \batitle#1{#1}\fi
\ifx \bjtitle  \undefined \def \bjtitle#1{#1}\fi
\ifx \bvolume  \undefined \def \bvolume#1{\textbf{#1}}\fi
\ifx \byear  \undefined \def \byear#1{#1}\fi
\ifx \bissue  \undefined \def \bissue#1{#1}\fi
\ifx \bfpage  \undefined \def \bfpage#1{#1}\fi
\ifx \blpage  \undefined \def \blpage #1{#1}\fi
\ifx \burl  \undefined \def \burl#1{\textsf{#1}}\fi
\ifx \doiurl  \undefined \def \doiurl#1{\url{https://doi.org/#1}}\fi
\ifx \betal  \undefined \def \betal{\textit{et al.}}\fi
\ifx \binstitute  \undefined \def \binstitute#1{#1}\fi
\ifx \binstitutionaled  \undefined \def \binstitutionaled#1{#1}\fi
\ifx \bctitle  \undefined \def \bctitle#1{#1}\fi
\ifx \beditor  \undefined \def \beditor#1{#1}\fi
\ifx \bpublisher  \undefined \def \bpublisher#1{#1}\fi
\ifx \bbtitle  \undefined \def \bbtitle#1{#1}\fi
\ifx \bedition  \undefined \def \bedition#1{#1}\fi
\ifx \bseriesno  \undefined \def \bseriesno#1{#1}\fi
\ifx \blocation  \undefined \def \blocation#1{#1}\fi
\ifx \bsertitle  \undefined \def \bsertitle#1{#1}\fi
\ifx \bsnm \undefined \def \bsnm#1{#1}\fi
\ifx \bsuffix \undefined \def \bsuffix#1{#1}\fi
\ifx \bparticle \undefined \def \bparticle#1{#1}\fi
\ifx \barticle \undefined \def \barticle#1{#1}\fi
\bibcommenthead
\ifx \bconfdate \undefined \def \bconfdate #1{#1}\fi
\ifx \botherref \undefined \def \botherref #1{#1}\fi
\ifx \url \undefined \def \url#1{\textsf{#1}}\fi
\ifx \bchapter \undefined \def \bchapter#1{#1}\fi
\ifx \bbook \undefined \def \bbook#1{#1}\fi
\ifx \bcomment \undefined \def \bcomment#1{#1}\fi
\ifx \oauthor \undefined \def \oauthor#1{#1}\fi
\ifx \citeauthoryear \undefined \def \citeauthoryear#1{#1}\fi
\ifx \endbibitem  \undefined \def \endbibitem {}\fi
\ifx \bconflocation  \undefined \def \bconflocation#1{#1}\fi
\ifx \arxivurl  \undefined \def \arxivurl#1{\textsf{#1}}\fi
\csname PreBibitemsHook\endcsname

%%% 1
\bibitem[\protect\citeauthoryear{Grelu and
  Akhmediev}{2012}]{grelu2012dissipative}
\begin{barticle}
\bauthor{\bsnm{Grelu}, \binits{P.}},
\bauthor{\bsnm{Akhmediev}, \binits{N.}}:
\batitle{Dissipative solitons for mode-locked lasers}.
\bjtitle{Nature photonics}
\bvolume{6}(\bissue{2}),
\bfpage{84}--\blpage{92}
(\byear{2012})
\end{barticle}
\endbibitem

%%% 2
\bibitem[\protect\citeauthoryear{Akhmediev and Ankiewicz}{2005}]{Akhmediev2005}
\begin{bbook}
\bauthor{\bsnm{Akhmediev}, \binits{N.}},
\bauthor{\bsnm{Ankiewicz}, \binits{A.}}:
In: \beditor{\bsnm{Akhmediev}, \binits{N.}},
\beditor{\bsnm{Ankiewicz}, \binits{A.}} (eds.)
\bbtitle{Dissipative Solitons in the Complex Ginzburg-Landau and
  Swift-Hohenberg Equations},
pp. \bfpage{1}--\blpage{17}.
\bpublisher{Springer},
\blocation{Berlin, Heidelberg}
(\byear{2005}).
\doiurl{10.1007/10928028_1} .
\burl{https://doi.org/10.1007/10928028_1}
\end{bbook}
\endbibitem

%%% 3
\bibitem[\protect\citeauthoryear{Ankiewicz and
  Akhmediev}{2008}]{ankiewicz2008dissipative}
\begin{bbook}
\bauthor{\bsnm{Ankiewicz}, \binits{A.}},
\bauthor{\bsnm{Akhmediev}, \binits{N.}}:
\bbtitle{Dissipative Solitons: from Optics to Biology and Medicine}.
\bpublisher{Springer},
\blocation{Heidelberg}
(\byear{2008})
\end{bbook}
\endbibitem

%%% 4
\bibitem[\protect\citeauthoryear{Purwins et~al.}{2010}]{purwins2010dissipative}
\begin{barticle}
\bauthor{\bsnm{Purwins}, \binits{H.-G.}},
\bauthor{\bsnm{B{\"o}deker}, \binits{H.}},
\bauthor{\bsnm{Amiranashvili}, \binits{S.}}:
\batitle{Dissipative solitons}.
\bjtitle{Advances in Physics}
\bvolume{59}(\bissue{5}),
\bfpage{485}--\blpage{701}
(\byear{2010})
\end{barticle}
\endbibitem

%%% 5
\bibitem[\protect\citeauthoryear{Ferreira}{2022}]{ferreira2022dissipative}
\begin{bbook}
\bauthor{\bsnm{Ferreira}, \binits{M.F.}}:
\bbtitle{Dissipative Optical Solitons}
vol. \bseriesno{238}.
\bpublisher{Springer}, \blocation{???}
(\byear{2022})
\end{bbook}
\endbibitem

%%% 6
\bibitem[\protect\citeauthoryear{Akhmediev and
  Ankiewicz}{2009}]{akhmediev2009dissipative}
\begin{barticle}
\bauthor{\bsnm{Akhmediev}, \binits{N.}},
\bauthor{\bsnm{Ankiewicz}, \binits{A.}}:
\batitle{Dissipative soliton resonances}.
\bjtitle{Physics Letters A}
\bvolume{373}(\bissue{17}),
\bfpage{2137}--\blpage{2145}
(\byear{2009})
\end{barticle}
\endbibitem

%%% 7
\bibitem[\protect\citeauthoryear{Grelu et~al.}{2010}]{grelu2010dissipative}
\begin{barticle}
\bauthor{\bsnm{Grelu}, \binits{P.}},
\bauthor{\bsnm{Chang}, \binits{W.}},
\bauthor{\bsnm{Ankiewicz}, \binits{A.}},
\bauthor{\bsnm{Soto-Crespo}, \binits{J.M.}},
\bauthor{\bsnm{Akhmediev}, \binits{N.}}:
\batitle{Dissipative soliton resonance as a guideline for high-energy pulse
  laser oscillators}.
\bjtitle{JOSA B}
\bvolume{27}(\bissue{11}),
\bfpage{2336}--\blpage{2341}
(\byear{2010})
\end{barticle}
\endbibitem

%%% 8
\bibitem[\protect\citeauthoryear{Malomed}{2005}]{malomed2005dissipative}
\begin{barticle}
\bauthor{\bsnm{Malomed}, \binits{B.A.}}:
\batitle{Dissipative solitons in bose-einstein condensates with a magnetic
  trap}.
\bjtitle{Physical Review A}
\bvolume{72}(\bissue{5}),
\bfpage{053610}
(\byear{2005})
\end{barticle}
\endbibitem

%%% 9
\bibitem[\protect\citeauthoryear{Fisher et~al.}{2009}]{fisher2009collective}
\begin{barticle}
\bauthor{\bsnm{Fisher}, \binits{B.M.}},
\bauthor{\bsnm{Lai}, \binits{C.K.}},
\bauthor{\bsnm{Malomed}, \binits{B.A.}}:
\batitle{Collective excitations and dissipative solitons in a strongly
  interacting bose-einstein condensate}.
\bjtitle{Physical Review Letters}
\bvolume{103}(\bissue{13}),
\bfpage{135301}
(\byear{2009})
\end{barticle}
\endbibitem

%%% 10
\bibitem[\protect\citeauthoryear{Chavanis}{2017}]{chavanis2017dissipative}
\begin{barticle}
\bauthor{\bsnm{Chavanis}, \binits{P.-H.}}:
\batitle{Dissipative self-gravitating bose-einstein condensates with arbitrary
  nonlinearity as a model of dark matter halos}.
\bjtitle{The European Physical Journal Plus}
\bvolume{132},
\bfpage{1}--\blpage{61}
(\byear{2017})
\end{barticle}
\endbibitem

%%% 11
\bibitem[\protect\citeauthoryear{Parker and
  Zhang}{1993}]{parker1993relativistic}
\begin{barticle}
\bauthor{\bsnm{Parker}, \binits{L.}},
\bauthor{\bsnm{Zhang}, \binits{Y.}}:
\batitle{Relativistic condensate as a source for inflation}.
\bjtitle{Physical Review D}
\bvolume{47}(\bissue{2}),
\bfpage{416}
(\byear{1993})
\end{barticle}
\endbibitem

%%% 12
\bibitem[\protect\citeauthoryear{Vespignani}{2012}]{vespignani2012modelling}
\begin{barticle}
\bauthor{\bsnm{Vespignani}, \binits{A.}}:
\batitle{Modelling dynamical processes in complex socio-technical systems}.
\bjtitle{Nature physics}
\bvolume{8}(\bissue{1}),
\bfpage{32}--\blpage{39}
(\byear{2012})
\end{barticle}
\endbibitem

%%% 13
\bibitem[\protect\citeauthoryear{Kalashnikov}{2016}]{kalashnikov2016optics}
\begin{bchapter}
\bauthor{\bsnm{Kalashnikov}, \binits{V.L.}}:
\bctitle{Optics and chaos: Chaotic, rogue, and noisy optical dissipative
  solitons}.
In: \beditor{\bsnm{Skiadas}, \binits{C.H.}},
\beditor{\bsnm{Ch.}, \binits{S.}} (eds.)
\bbtitle{Handbook of Applications of Chaos Theory},
pp. \bfpage{587}--\blpage{626}.
\bpublisher{CRC Press},
\blocation{New {Y}ork}
(\byear{2016})
\end{bchapter}
\endbibitem

%%% 14
\bibitem[\protect\citeauthoryear{Picozzi et~al.}{2014}]{picozzi2014optical}
\begin{barticle}
\bauthor{\bsnm{Picozzi}, \binits{A.}},
\bauthor{\bsnm{Garnier}, \binits{J.}},
\bauthor{\bsnm{Hansson}, \binits{T.}},
\bauthor{\bsnm{Suret}, \binits{P.}},
\bauthor{\bsnm{Randoux}, \binits{S.}},
\bauthor{\bsnm{Millot}, \binits{G.}},
\bauthor{\bsnm{Christodoulides}, \binits{D.N.}}:
\batitle{Optical wave turbulence: Towards a unified nonequilibrium
  thermodynamic formulation of statistical nonlinear optics}.
\bjtitle{Physics Reports}
\bvolume{542}(\bissue{1}),
\bfpage{1}--\blpage{132}
(\byear{2014})
\end{barticle}
\endbibitem

%%% 15
\bibitem[\protect\citeauthoryear{Kalashnikov
  et~al.}{2023}]{kalashnikov2023dissipative}
\begin{botherref}
\oauthor{\bsnm{Kalashnikov}, \binits{V.L.}},
\oauthor{\bsnm{Rudenkov}, \binits{A.}},
\oauthor{\bsnm{Sorokin}, \binits{E.}},
\oauthor{\bsnm{Sorokina}, \binits{I.}}:
Dissipative soliton resonance: Adiabatic theory and thermodynamics.
arXiv preprint arXiv:2305.00516
(2023)
\end{botherref}
\endbibitem

%%% 16
\bibitem[\protect\citeauthoryear{Churkin et~al.}{2015}]{churkin2015wave}
\begin{barticle}
\bauthor{\bsnm{Churkin}, \binits{D.V.}},
\bauthor{\bsnm{Kolokolov}, \binits{I.V.}},
\bauthor{\bsnm{Podivilov}, \binits{E.V.}},
\bauthor{\bsnm{Vatnik}, \binits{I.D.}},
\bauthor{\bsnm{Nikulin}, \binits{M.A.}},
\bauthor{\bsnm{Vergeles}, \binits{S.S.}},
\bauthor{\bsnm{Terekhov}, \binits{I.S.}},
\bauthor{\bsnm{Lebedev}, \binits{V.V.}},
\bauthor{\bsnm{Falkovich}, \binits{G.}},
\bauthor{\bsnm{Babin}, \binits{S.A.}}, \betal:
\batitle{Wave kinetics of random fibre lasers}.
\bjtitle{Nature Communications}
\bvolume{6}(\bissue{1}),
\bfpage{6214}
(\byear{2015})
\end{barticle}
\endbibitem

%%% 17
\bibitem[\protect\citeauthoryear{Baudin et~al.}{2023}]{baudin2023observation}
\begin{barticle}
\bauthor{\bsnm{Baudin}, \binits{K.}},
\bauthor{\bsnm{Garnier}, \binits{J.}},
\bauthor{\bsnm{Fusaro}, \binits{A.}},
\bauthor{\bsnm{Berti}, \binits{N.}},
\bauthor{\bsnm{Michel}, \binits{C.}},
\bauthor{\bsnm{Krupa}, \binits{K.}},
\bauthor{\bsnm{Millot}, \binits{G.}},
\bauthor{\bsnm{Picozzi}, \binits{A.}}:
\batitle{Observation of light thermalization to negative-temperature
  rayleigh-jeans equilibrium states in multimode optical fibers}.
\bjtitle{Physical Review Letters}
\bvolume{130}(\bissue{6}),
\bfpage{063801}
(\byear{2023})
\end{barticle}
\endbibitem

%%% 18
\bibitem[\protect\citeauthoryear{Wu et~al.}{2019}]{wu2019thermodynamic}
\begin{barticle}
\bauthor{\bsnm{Wu}, \binits{F.O.}},
\bauthor{\bsnm{Tikan}, \binits{A.}},
\bauthor{\bsnm{Karpov}, \binits{M.}},
\bauthor{\bsnm{Liu}, \binits{J.}},
\bauthor{\bsnm{Guo}, \binits{H.}},
\bauthor{\bsnm{Chen}, \binits{Y.}},
\bauthor{\bsnm{Yan}, \binits{Y.}},
\bauthor{\bsnm{Mikkelsen}, \binits{M.H.}},
\bauthor{\bsnm{Liu}, \binits{D.}},
\bauthor{\bsnm{Chen}, \binits{L.}},
\bauthor{\bsnm{al.}}:
\batitle{Thermodynamic theory of highly multimoded nonlinear optical systems}.
\bjtitle{Nature Photonics}
\bvolume{13}(\bissue{11}),
\bfpage{776}--\blpage{783}
(\byear{2019})
\end{barticle}
\endbibitem

%%% 19
\bibitem[\protect\citeauthoryear{Zakharov et~al.}{2004}]{zakharov2004one}
\begin{barticle}
\bauthor{\bsnm{Zakharov}, \binits{V.}},
\bauthor{\bsnm{Dias}, \binits{F.}},
\bauthor{\bsnm{Pushkarev}, \binits{A.}}:
\batitle{One-dimensional wave turbulence}.
\bjtitle{Physics Reports}
\bvolume{398}(\bissue{1}),
\bfpage{1}--\blpage{65}
(\byear{2004})
\end{barticle}
\endbibitem

%%% 20
\bibitem[\protect\citeauthoryear{Picozzi
  et~al.}{2009}]{picozzi2009thermalization}
\begin{barticle}
\bauthor{\bsnm{Picozzi}, \binits{A.}},
\bauthor{\bsnm{Barviau}, \binits{B.}},
\bauthor{\bsnm{Kibler}, \binits{B.}},
\bauthor{\bsnm{Rica}, \binits{S.}}:
\batitle{Thermalization of incoherent nonlinear waves: From incoherent solitons
  to a thermodynamic description of statistical nonlinear optics}.
\bjtitle{The European Physical Journal special topics}
\bvolume{173}(\bissue{1}),
\bfpage{313}--\blpage{340}
(\byear{2009})
\end{barticle}
\endbibitem

%%% 21
\bibitem[\protect\citeauthoryear{Dyachenko et~al.}{1992}]{dyachenko1992optical}
\begin{barticle}
\bauthor{\bsnm{Dyachenko}, \binits{S.}},
\bauthor{\bsnm{Newell}, \binits{A.}},
\bauthor{\bsnm{Pushkarev}, \binits{A.}},
\bauthor{\bsnm{Zakharov}, \binits{V.}}:
\batitle{Optical turbulence: weak turbulence, condensates and collapsing
  filaments in the nonlinear schr{\"o}dinger equation}.
\bjtitle{Physica D: Nonlinear Phenomena}
\bvolume{57}(\bissue{1-2}),
\bfpage{96}--\blpage{160}
(\byear{1992})
\end{barticle}
\endbibitem

%%% 22
\bibitem[\protect\citeauthoryear{Robinson}{1997}]{robinson1997nonlinear}
\begin{barticle}
\bauthor{\bsnm{Robinson}, \binits{P.}}:
\batitle{Nonlinear wave collapse and strong turbulence}.
\bjtitle{Reviews of modern physics}
\bvolume{69}(\bissue{2}),
\bfpage{507}
(\byear{1997})
\end{barticle}
\endbibitem

%%% 23
\bibitem[\protect\citeauthoryear{Picozzi}{2007}]{picozzi2007towards}
\begin{barticle}
\bauthor{\bsnm{Picozzi}, \binits{A.}}:
\batitle{Towards a nonequilibrium thermodynamic description of incoherent
  nonlinear optics}.
\bjtitle{Optics Express}
\bvolume{15}(\bissue{14}),
\bfpage{9063}--\blpage{9083}
(\byear{2007})
\end{barticle}
\endbibitem

%%% 24
\bibitem[\protect\citeauthoryear{Nazarenko}{2011}]{nazarenko2011wave}
\begin{bbook}
\bauthor{\bsnm{Nazarenko}, \binits{S.}}:
\bbtitle{Wave Turbulence}
vol. \bseriesno{825}.
\bpublisher{Springer},
\blocation{Berlin}
(\byear{2011})
\end{bbook}
\endbibitem

%%% 25
\bibitem[\protect\citeauthoryear{Marcuvitz}{1980}]{marcuvitz1980quasiparticle}
\begin{barticle}
\bauthor{\bsnm{Marcuvitz}, \binits{N.}}:
\batitle{Quasiparticle view of wave propagation}.
\bjtitle{Proceedings of the IEEE}
\bvolume{68}(\bissue{11}),
\bfpage{1380}--\blpage{1395}
(\byear{1980})
\end{barticle}
\endbibitem

%%% 26
\bibitem[\protect\citeauthoryear{Akhmediev
  et~al.}{1998}]{akhmediev1998partially}
\begin{barticle}
\bauthor{\bsnm{Akhmediev}, \binits{N.}},
\bauthor{\bsnm{Kr{\'o}likowski}, \binits{W.}},
\bauthor{\bsnm{Snyder}, \binits{A.}}:
\batitle{Partially coherent solitons of variable shape}.
\bjtitle{Physical Review Letters}
\bvolume{81}(\bissue{21}),
\bfpage{4632}
(\byear{1998})
\end{barticle}
\endbibitem

%%% 27
\bibitem[\protect\citeauthoryear{Sob'yanin}{2013}]{sob2013bose}
\begin{barticle}
\bauthor{\bsnm{Sob'yanin}, \binits{D.N.}}:
\batitle{Bose-einstein condensation of light: General theory}.
\bjtitle{Physical Review E}
\bvolume{88}(\bissue{2}),
\bfpage{022132}
(\byear{2013})
\end{barticle}
\endbibitem

%%% 28
\bibitem[\protect\citeauthoryear{Kalashnikov and
  Wabnitz}{2021}]{kalashnikov2021metaphorical}
\begin{barticle}
\bauthor{\bsnm{Kalashnikov}, \binits{V.L.}},
\bauthor{\bsnm{Wabnitz}, \binits{S.}}:
\batitle{A “metaphorical” nonlinear multimode fiber laser approach to
  weakly dissipative bose-einstein condensates}.
\bjtitle{Europhysics Letters}
\bvolume{133}(\bissue{3}),
\bfpage{34002}
(\byear{2021})
\end{barticle}
\endbibitem

%%% 29
\bibitem[\protect\citeauthoryear{Carusotto and
  Ciuti}{2013}]{carusotto2013quantum}
\begin{barticle}
\bauthor{\bsnm{Carusotto}, \binits{I.}},
\bauthor{\bsnm{Ciuti}, \binits{C.}}:
\batitle{Quantum fluids of light}.
\bjtitle{Reviews of Modern Physics}
\bvolume{85}(\bissue{1}),
\bfpage{299}
(\byear{2013})
\end{barticle}
\endbibitem

%%% 30
\bibitem[\protect\citeauthoryear{Gordon and
  Fischer}{2003}]{gordon2003inhibition}
\begin{barticle}
\bauthor{\bsnm{Gordon}, \binits{A.}},
\bauthor{\bsnm{Fischer}, \binits{B.}}:
\batitle{Inhibition of modulation instability in lasers by noise}.
\bjtitle{Optics letters}
\bvolume{28}(\bissue{15}),
\bfpage{1326}--\blpage{1328}
(\byear{2003})
\end{barticle}
\endbibitem

%%% 31
\bibitem[\protect\citeauthoryear{Gat et~al.}{2004}]{gat2004solution}
\begin{barticle}
\bauthor{\bsnm{Gat}, \binits{O.}},
\bauthor{\bsnm{Gordon}, \binits{A.}},
\bauthor{\bsnm{Fischer}, \binits{B.}}:
\batitle{Solution of a statistical mechanics model for pulse formation in
  lasers}.
\bjtitle{Physical Review E}
\bvolume{70}(\bissue{4}),
\bfpage{046108}
(\byear{2004})
\end{barticle}
\endbibitem

%%% 32
\bibitem[\protect\citeauthoryear{Gordon et~al.}{2006}]{gordon2006self}
\begin{barticle}
\bauthor{\bsnm{Gordon}, \binits{A.}},
\bauthor{\bsnm{Gat}, \binits{O.}},
\bauthor{\bsnm{Fischer}, \binits{B.}},
\bauthor{\bsnm{K{\"a}rtner}, \binits{F.X.}}:
\batitle{Self-starting of passive mode locking}.
\bjtitle{Optics Express}
\bvolume{14}(\bissue{23}),
\bfpage{11142}--\blpage{11154}
(\byear{2006})
\end{barticle}
\endbibitem

%%% 33
\bibitem[\protect\citeauthoryear{Gordon and Fischer}{2002}]{gordon2002phase}
\begin{barticle}
\bauthor{\bsnm{Gordon}, \binits{A.}},
\bauthor{\bsnm{Fischer}, \binits{B.}}:
\batitle{Phase transition theory of many-mode ordering and pulse formation in
  lasers}.
\bjtitle{Physical review letters}
\bvolume{89}(\bissue{10}),
\bfpage{103901}
(\byear{2002})
\end{barticle}
\endbibitem

%%% 34
\bibitem[\protect\citeauthoryear{Werner and Friberg}{1997}]{werner1997phase}
\begin{barticle}
\bauthor{\bsnm{Werner}, \binits{M.}},
\bauthor{\bsnm{Friberg}, \binits{S.}}:
\batitle{Phase transitions and the internal noise structure of nonlinear
  schr{\"o}dinger equation solitons}.
\bjtitle{Physical review letters}
\bvolume{79}(\bissue{21}),
\bfpage{4143}
(\byear{1997})
\end{barticle}
\endbibitem

%%% 35
\bibitem[\protect\citeauthoryear{Kalashnikov and
  Sorokin}{2018}]{kalashnikov2018self}
\begin{bchapter}
\bauthor{\bsnm{Kalashnikov}, \binits{V.L.}},
\bauthor{\bsnm{Sorokin}, \binits{E.}}:
\bctitle{Self-organization, coherence and turbulence in laser optics}.
In: \beditor{\bsnm{Lopez-Ruiz}, \binits{R.}} (ed.)
\bbtitle{Complexity in Biological and Physical Systems},
pp. \bfpage{97}--\blpage{112}.
\bpublisher{IntechOpen},
\blocation{Rijeka}
(\byear{2018})
\end{bchapter}
\endbibitem

%%% 36
\bibitem[\protect\citeauthoryear{Malomed}{2006}]{malomed1}
\begin{bchapter}
\bauthor{\bsnm{Malomed}, \binits{B.}}:
\bctitle{Nonlinear schr{\"o}dinger equations}.
In: \beditor{\bsnm{Scott}, \binits{A.}} (ed.)
\bbtitle{Encyclopedia of Nonlinear Science},
pp. \bfpage{639}--\blpage{643}.
\bpublisher{Routledge},
\blocation{New {Y}ork}
(\byear{2006})
\end{bchapter}
\endbibitem

%%% 37
\bibitem[\protect\citeauthoryear{Aranson and Kramer}{2002}]{aranson2002world}
\begin{barticle}
\bauthor{\bsnm{Aranson}, \binits{I.S.}},
\bauthor{\bsnm{Kramer}, \binits{L.}}:
\batitle{The world of the complex ginzburg-landau equation}.
\bjtitle{Reviews of modern physics}
\bvolume{74}(\bissue{1}),
\bfpage{99}
(\byear{2002})
\end{barticle}
\endbibitem

%%% 38
\bibitem[\protect\citeauthoryear{Podivilov and
  Kalashnikov}{2005}]{podivilov2005heavily}
\begin{barticle}
\bauthor{\bsnm{Podivilov}, \binits{E.}},
\bauthor{\bsnm{Kalashnikov}, \binits{V.L.}}:
\batitle{Heavily-chirped solitary pulses in the normal dispersion region: new
  solutions of the cubic-quintic complex ginzburg-landau equation}.
\bjtitle{Journal of Experimental and Theoretical Physics Letters}
\bvolume{82},
\bfpage{467}--\blpage{471}
(\byear{2005})
\end{barticle}
\endbibitem

%%% 39
\bibitem[\protect\citeauthoryear{Katz et~al.}{2006}]{katz2006non}
\begin{barticle}
\bauthor{\bsnm{Katz}, \binits{M.}},
\bauthor{\bsnm{Gordon}, \binits{A.}},
\bauthor{\bsnm{Gat}, \binits{O.}},
\bauthor{\bsnm{Fischer}, \binits{B.}}:
\batitle{Non-gibbsian stochastic light-mode dynamics of passive mode locking}.
\bjtitle{Physical review letters}
\bvolume{97}(\bissue{11}),
\bfpage{113902}
(\byear{2006})
\end{barticle}
\endbibitem

%%% 40
\bibitem[\protect\citeauthoryear{Chang et~al.}{2008}]{chang2008dissipative}
\begin{barticle}
\bauthor{\bsnm{Chang}, \binits{W.}},
\bauthor{\bsnm{Ankiewicz}, \binits{A.}},
\bauthor{\bsnm{Soto-Crespo}, \binits{J.}},
\bauthor{\bsnm{Akhmediev}, \binits{N.}}:
\batitle{Dissipative soliton resonances}.
\bjtitle{Physical Review A}
\bvolume{78}(\bissue{2}),
\bfpage{023830}
(\byear{2008})
\end{barticle}
\endbibitem

%%% 41
\bibitem[\protect\citeauthoryear{Kharenko et~al.}{2011}]{kharenko2011highly}
\begin{barticle}
\bauthor{\bsnm{Kharenko}, \binits{D.S.}},
\bauthor{\bsnm{Shtyrina}, \binits{O.V.}},
\bauthor{\bsnm{Yarutkina}, \binits{I.A.}},
\bauthor{\bsnm{Podivilov}, \binits{E.V.}},
\bauthor{\bsnm{Fedoruk}, \binits{M.P.}},
\bauthor{\bsnm{Babin}, \binits{S.A.}}:
\batitle{Highly chirped dissipative solitons as a one-parameter family of
  stable solutions of the cubic--quintic ginzburg--landau equation}.
\bjtitle{JOSA B}
\bvolume{28}(\bissue{10}),
\bfpage{2314}--\blpage{2319}
(\byear{2011})
\end{barticle}
\endbibitem

%%% 42
\bibitem[\protect\citeauthoryear{Akhmanov et~al.}{1992}]{akhmanov1992optics}
\begin{bbook}
\bauthor{\bsnm{Akhmanov}, \binits{S.A.}},
\bauthor{\bsnm{Vysloukh}, \binits{V.A.}},
\bauthor{\bsnm{Chirkin}, \binits{A.S.}},
\bauthor{\bsnm{Atanov}, \binits{Y.}}:
\bbtitle{Optics of Femtosecond Laser Pulses}.
\bpublisher{Springer},
\blocation{Melville, NY}
(\byear{1992})
\end{bbook}
\endbibitem

%%% 43
\bibitem[\protect\citeauthoryear{Laurie et~al.}{2012}]{laurie2012one}
\begin{barticle}
\bauthor{\bsnm{Laurie}, \binits{J.}},
\bauthor{\bsnm{Bortolozzo}, \binits{U.}},
\bauthor{\bsnm{Nazarenko}, \binits{S.}},
\bauthor{\bsnm{Residori}, \binits{S.}}:
\batitle{One-dimensional optical wave turbulence: experiment and theory}.
\bjtitle{Physics Reports}
\bvolume{514}(\bissue{4}),
\bfpage{121}--\blpage{175}
(\byear{2012})
\end{barticle}
\endbibitem

%%% 44
\bibitem[\protect\citeauthoryear{Sorokin et~al.}{2013}]{sorokin2013chaotic}
\begin{barticle}
\bauthor{\bsnm{Sorokin}, \binits{E.}},
\bauthor{\bsnm{Tolstik}, \binits{N.}},
\bauthor{\bsnm{Kalashnikov}, \binits{V.L.}},
\bauthor{\bsnm{Sorokina}, \binits{I.T.}}:
\batitle{Chaotic chirped-pulse oscillators}.
\bjtitle{Optics Express}
\bvolume{21}(\bissue{24}),
\bfpage{29567}--\blpage{29577}
(\byear{2013})
\end{barticle}
\endbibitem

%%% 45
\bibitem[\protect\citeauthoryear{Kalashnikov}{2009}]{maple}
\begin{botherref}
\oauthor{\bsnm{Kalashnikov}, \binits{V.L.}}:
Chirped solitary-pulse solutions of the complex cubic-quintic nonlinear
  Ginzburg-Landau equation.
Maple worksheet
  \url{https://www.researchgate.net/publication/284730587_Chirped_solitary-pulse_solutions_of_the_complex_cubic-quintic_nonlinear_Ginzburg-Landau_equation}
(2009)
\end{botherref}
\endbibitem

%%% 46
\bibitem[\protect\citeauthoryear{Zhu et~al.}{2013}]{zhu2013generation}
\begin{barticle}
\bauthor{\bsnm{Zhu}, \binits{L.}},
\bauthor{\bsnm{Verhoef}, \binits{A.}},
\bauthor{\bsnm{Jespersen}, \binits{K.}},
\bauthor{\bsnm{Kalashnikov}, \binits{V.}},
\bauthor{\bsnm{Gr{\"u}ner-Nielsen}, \binits{L.}},
\bauthor{\bsnm{Lorenc}, \binits{D.}},
\bauthor{\bsnm{Baltu{\v{s}}ka}, \binits{A.}},
\bauthor{\bsnm{Fern{\'a}ndez}, \binits{A.}}:
\batitle{Generation of high fidelity 62-fs, 7-nj pulses at 1035 nm from a net
  normal-dispersion yb-fiber laser with anomalous dispersion higher-order-mode
  fiber}.
\bjtitle{Optics express}
\bvolume{21}(\bissue{14}),
\bfpage{16255}--\blpage{16262}
(\byear{2013})
\end{barticle}
\endbibitem

%%% 47
\bibitem[\protect\citeauthoryear{Rudenkov et~al.}{2023}]{rudenkov2023high}
\begin{barticle}
\bauthor{\bsnm{Rudenkov}, \binits{A.}},
\bauthor{\bsnm{Kalashnikov}, \binits{V.L.}},
\bauthor{\bsnm{Sorokin}, \binits{E.}},
\bauthor{\bsnm{Demesh}, \binits{M.}},
\bauthor{\bsnm{Sorokina}, \binits{I.T.}}:
\batitle{High peak power and energy scaling in the mid-ir chirped-pulse
  oscillator-amplifier laser systems}.
\bjtitle{Optics Express}
\bvolume{31}(\bissue{11}),
\bfpage{17820}--\blpage{17835}
(\byear{2023})
\end{barticle}
\endbibitem

%%% 48
\bibitem[\protect\citeauthoryear{Kalashnikov
  et~al.}{2005}]{kalashnikov2005approaching}
\begin{barticle}
\bauthor{\bsnm{Kalashnikov}, \binits{V.L.}},
\bauthor{\bsnm{Podivilov}, \binits{E.}},
\bauthor{\bsnm{Chernykh}, \binits{A.}},
\bauthor{\bsnm{Naumov}, \binits{S.}},
\bauthor{\bsnm{Fernandez}, \binits{A.}},
\bauthor{\bsnm{Graf}, \binits{R.}},
\bauthor{\bsnm{Apolonski}, \binits{A.}}:
\batitle{Approaching the microjoule frontier with femtosecond laser
  oscillators: theory and comparison with experiment}.
\bjtitle{New Journal of Physics}
\bvolume{7}(\bissue{1}),
\bfpage{217}
(\byear{2005})
\end{barticle}
\endbibitem

%%% 49
\bibitem[\protect\citeauthoryear{Bale et~al.}{2008}]{bale2008spectral}
\begin{barticle}
\bauthor{\bsnm{Bale}, \binits{B.G.}},
\bauthor{\bsnm{Kutz}, \binits{J.N.}},
\bauthor{\bsnm{Chong}, \binits{A.}},
\bauthor{\bsnm{Renninger}, \binits{W.H.}},
\bauthor{\bsnm{Wise}, \binits{F.W.}}:
\batitle{Spectral filtering for high-energy mode-locking in normal dispersion
  fiber lasers}.
\bjtitle{JOSA B}
\bvolume{25}(\bissue{10}),
\bfpage{1763}--\blpage{1770}
(\byear{2008})
\end{barticle}
\endbibitem

%%% 50
\bibitem[\protect\citeauthoryear{Kalashnikov}{2023}]{maple2}
\begin{botherref}
\oauthor{\bsnm{Kalashnikov}, \binits{V.L.}}:
Strongly Chirped Dissipative Soliton of the Complex Nonlinear Cubic-quintic
  Ginzburg-Landau Equation Without a Spectral Dissipation,
ResearchGate
(2023).
Maple worksheet \url{http://dx.doi.org/10.13140/RG.2.2.35780.40324}
\end{botherref}
\endbibitem

%%% 51
\bibitem[\protect\citeauthoryear{Kalashnikov and
  Apolonski}{2009}]{PhysRevA.79.043829}
\begin{barticle}
\bauthor{\bsnm{Kalashnikov}, \binits{V.L.}},
\bauthor{\bsnm{Apolonski}, \binits{A.}}:
\batitle{Chirped-pulse oscillators: A unified standpoint}.
\bjtitle{Phys. Rev. A}
\bvolume{79},
\bfpage{043829}
(\byear{2009})
\doiurl{10.1103/PhysRevA.79.043829}
\end{barticle}
\endbibitem

%%% 52
\bibitem[\protect\citeauthoryear{Kalashnikov
  et~al.}{2006}]{kalashnikov2006chirped}
\begin{barticle}
\bauthor{\bsnm{Kalashnikov}, \binits{V.L.}},
\bauthor{\bsnm{Podivilov}, \binits{E.}},
\bauthor{\bsnm{Chernykh}, \binits{A.}},
\bauthor{\bsnm{Apolonski}, \binits{A.}}:
\batitle{Chirped-pulse oscillators: theory and experiment}.
\bjtitle{Applied Physics B}
\bvolume{83},
\bfpage{503}--\blpage{510}
(\byear{2006})
\end{barticle}
\endbibitem

%%% 53
\bibitem[\protect\citeauthoryear{Haus and Mecozzi}{1993}]{206583}
\begin{barticle}
\bauthor{\bsnm{Haus}, \binits{H.A.}},
\bauthor{\bsnm{Mecozzi}, \binits{A.}}:
\batitle{Noise of mode-locked lasers}.
\bjtitle{IEEE Journal of Quantum Electronics}
\bvolume{29}(\bissue{3}),
\bfpage{983}--\blpage{996}
(\byear{1993})
\doiurl{10.1109/3.206583}
\end{barticle}
\endbibitem

%%% 54
\bibitem[\protect\citeauthoryear{Gardiner and
  Zoller}{2004}]{gardiner2004quantum}
\begin{bbook}
\bauthor{\bsnm{Gardiner}, \binits{C.}},
\bauthor{\bsnm{Zoller}, \binits{P.}}:
\bbtitle{Quantum Noise: a Handbook of Markovian and non-Markovian Quantum
  Stochastic Methods with Applications to Quantum Optics}.
\bpublisher{Springer}, \blocation{???}
(\byear{2004})
\end{bbook}
\endbibitem

%%% 55
\bibitem[\protect\citeauthoryear{Hall et~al.}{2002}]{hall2002statistical}
\begin{barticle}
\bauthor{\bsnm{Hall}, \binits{B.}},
\bauthor{\bsnm{Lisak}, \binits{M.}},
\bauthor{\bsnm{Anderson}, \binits{D.}},
\bauthor{\bsnm{Fedele}, \binits{R.}},
\bauthor{\bsnm{Semenov}, \binits{V.}}:
\batitle{Statistical theory for incoherent light propagation in nonlinear
  media}.
\bjtitle{Physical Review E}
\bvolume{65}(\bissue{3}),
\bfpage{035602}
(\byear{2002})
\end{barticle}
\endbibitem

%%% 56
\bibitem[\protect\citeauthoryear{D{\"u}ring et~al.}{2009}]{during2009breakdown}
\begin{barticle}
\bauthor{\bsnm{D{\"u}ring}, \binits{G.}},
\bauthor{\bsnm{Picozzi}, \binits{A.}},
\bauthor{\bsnm{Rica}, \binits{S.}}:
\batitle{Breakdown of weak-turbulence and nonlinear wave condensation}.
\bjtitle{Physica D: Nonlinear Phenomena}
\bvolume{238}(\bissue{16}),
\bfpage{1524}--\blpage{1549}
(\byear{2009})
\end{barticle}
\endbibitem

%%% 57
\bibitem[\protect\citeauthoryear{Krausz et~al.}{1991}]{krausz1991self}
\begin{barticle}
\bauthor{\bsnm{Krausz}, \binits{F.}},
\bauthor{\bsnm{Brabec}, \binits{T.}},
\bauthor{\bsnm{Spielmann}, \binits{C.}}:
\batitle{Self-starting passive mode locking}.
\bjtitle{Optics letters}
\bvolume{16}(\bissue{4}),
\bfpage{235}--\blpage{237}
(\byear{1991})
\end{barticle}
\endbibitem

%%% 58
\bibitem[\protect\citeauthoryear{Krausz and Brabec}{1993}]{krausz1993passive}
\begin{barticle}
\bauthor{\bsnm{Krausz}, \binits{F.}},
\bauthor{\bsnm{Brabec}, \binits{T.}}:
\batitle{Passive mode locking in standing-wave laser resonators}.
\bjtitle{Optics letters}
\bvolume{18}(\bissue{11}),
\bfpage{888}--\blpage{890}
(\byear{1993})
\end{barticle}
\endbibitem

%%% 59
\bibitem[\protect\citeauthoryear{Lugiato et~al.}{2015}]{lugiato2015nonlinear}
\begin{bbook}
\bauthor{\bsnm{Lugiato}, \binits{L.}},
\bauthor{\bsnm{Prati}, \binits{F.}},
\bauthor{\bsnm{Brambilla}, \binits{M.}}:
\bbtitle{Nonlinear Optical Systems}.
\bpublisher{Cambridge University Press}, \blocation{???}
(\byear{2015})
\end{bbook}
\endbibitem

%%% 60
\bibitem[\protect\citeauthoryear{Sergeyev et~al.}{2021}]{sergeyev2021vector}
\begin{barticle}
\bauthor{\bsnm{Sergeyev}, \binits{S.}},
\bauthor{\bsnm{Kolpakov}, \binits{S.}},
\bauthor{\bsnm{Loika}, \binits{Y.}}:
\batitle{Vector harmonic mode-locking by acoustic resonance}.
\bjtitle{Photonics Research}
\bvolume{9}(\bissue{8}),
\bfpage{1432}--\blpage{1438}
(\byear{2021})
\end{barticle}
\endbibitem

%%% 61
\bibitem[\protect\citeauthoryear{Ippen et~al.}{1990}]{ippen1990self}
\begin{barticle}
\bauthor{\bsnm{Ippen}, \binits{E.}},
\bauthor{\bsnm{Liu}, \binits{L.}},
\bauthor{\bsnm{Haus}, \binits{H.}}:
\batitle{Self-starting condition for additive-pulse mode-locked lasers}.
\bjtitle{Optics letters}
\bvolume{15}(\bissue{3}),
\bfpage{183}--\blpage{185}
(\byear{1990})
\end{barticle}
\endbibitem

%%% 62
\bibitem[\protect\citeauthoryear{Chen et~al.}{1995}]{chen1995self}
\begin{barticle}
\bauthor{\bsnm{Chen}, \binits{C.-J.}},
\bauthor{\bsnm{Wai}, \binits{P.K.A.}},
\bauthor{\bsnm{Menyuk}, \binits{C.}}:
\batitle{Self-starting of passively mode-locked lasers with fast saturable
  absorbers}.
\bjtitle{Optics letters}
\bvolume{20}(\bissue{4}),
\bfpage{350}--\blpage{352}
(\byear{1995})
\end{barticle}
\endbibitem

%%% 63
\bibitem[\protect\citeauthoryear{Soto-Crespo et~al.}{2002}]{soto2002continuous}
\begin{barticle}
\bauthor{\bsnm{Soto-Crespo}, \binits{J.M.}},
\bauthor{\bsnm{Akhmediev}, \binits{N.}},
\bauthor{\bsnm{Town}, \binits{G.}}:
\batitle{Continuous-wave versus pulse regime in a passively mode-locked laser
  with a fast saturable absorber}.
\bjtitle{JOSA B}
\bvolume{19}(\bissue{2}),
\bfpage{234}--\blpage{242}
(\byear{2002})
\end{barticle}
\endbibitem

%%% 64
\bibitem[\protect\citeauthoryear{Gordon and Fischer}{2003}]{gordon2003phase}
\begin{barticle}
\bauthor{\bsnm{Gordon}, \binits{A.}},
\bauthor{\bsnm{Fischer}, \binits{B.}}:
\batitle{Phase transition theory of pulse formation in passively mode-locked
  lasers with dispersion and kerr nonlinearity}.
\bjtitle{Optics communications}
\bvolume{223}(\bissue{1-3}),
\bfpage{151}--\blpage{156}
(\byear{2003})
\end{barticle}
\endbibitem

%%% 65
\bibitem[\protect\citeauthoryear{Lindner et~al.}{2004}]{lindner2004effects}
\begin{barticle}
\bauthor{\bsnm{Lindner}, \binits{B.}},
\bauthor{\bsnm{Garc{\i}a-Ojalvo}, \binits{J.}},
\bauthor{\bsnm{Neiman}, \binits{A.}},
\bauthor{\bsnm{Schimansky-Geier}, \binits{L.}}:
\batitle{Effects of noise in excitable systems}.
\bjtitle{Physics reports}
\bvolume{392}(\bissue{6}),
\bfpage{321}--\blpage{424}
(\byear{2004})
\end{barticle}
\endbibitem

%%% 66
\bibitem[\protect\citeauthoryear{Dudley et~al.}{2019}]{dudley2019rogue}
\begin{barticle}
\bauthor{\bsnm{Dudley}, \binits{J.M.}},
\bauthor{\bsnm{Genty}, \binits{G.}},
\bauthor{\bsnm{Mussot}, \binits{A.}},
\bauthor{\bsnm{Chabchoub}, \binits{A.}},
\bauthor{\bsnm{Dias}, \binits{F.}}:
\batitle{Rogue waves and analogies in optics and oceanography}.
\bjtitle{Nature Reviews Physics}
\bvolume{1}(\bissue{11}),
\bfpage{675}--\blpage{689}
(\byear{2019})
\end{barticle}
\endbibitem

%%% 67
\bibitem[\protect\citeauthoryear{Gordon et~al.}{2003}]{gordon2003melting}
\begin{barticle}
\bauthor{\bsnm{Gordon}, \binits{A.}},
\bauthor{\bsnm{Vodonos}, \binits{B.}},
\bauthor{\bsnm{Smulakovski}, \binits{V.}},
\bauthor{\bsnm{Fischer}, \binits{B.}}:
\batitle{Melting and freezing of light pulses and modes in mode-locked lasers}.
\bjtitle{Optics express}
\bvolume{11}(\bissue{25}),
\bfpage{3418}--\blpage{3424}
(\byear{2003})
\end{barticle}
\endbibitem

%%% 68
\bibitem[\protect\citeauthoryear{Hanggi}{1986}]{hanggi1986escape}
\begin{barticle}
\bauthor{\bsnm{Hanggi}, \binits{P.}}:
\batitle{Escape from a metastable state}.
\bjtitle{Journal of Statistical Physics}
\bvolume{42},
\bfpage{105}--\blpage{148}
(\byear{1986})
\end{barticle}
\endbibitem

%%% 69
\bibitem[\protect\citeauthoryear{Sj\"alander et~al.}{2019}]{sjalander}
\begin{botherref}
\oauthor{\bsnm{Sj\"alander}, \binits{M.}},
\oauthor{\bsnm{Jahre}, \binits{M.}},
\oauthor{\bsnm{Tufte}, \binits{G.}},
\oauthor{\bsnm{Reissmann}, \binits{N.}}:
{EPIC}: An Energy-Efficient, High-Performance {GPGPU} Computing Research
  Infrastructure
(2019)
\end{botherref}
\endbibitem

\end{thebibliography}

\end{document}